\documentclass[twocolumn]{aastex62}

\newcommand{\Sp}{{Spitzer\/}}

\usepackage{amsmath}
\usepackage{booktabs}

\begin{document}

\title{A Semi-automated Computational Approach for Infrared Dark Cloud Localization:\\ A Catalog of Infrared Dark Clouds}

\author{Jyothish Pari}
\affiliation{Center for Astrophysics $|$ Harvard \& Smithsonian, 60 Garden Street, MS-65, Cambridge, MA 02138-1516; USA}

\author[0000-0002-5599-4650]{Joseph L. Hora}
\affiliation{Center for Astrophysics $|$ Harvard \& Smithsonian, 60 Garden Street, MS-65, Cambridge, MA 02138-1516; USA}
\correspondingauthor{Joseph L. Hora}
\email{jhora@cfa.harvard.edu}

\accepted{to the Publications of the Astronomical Society of the Pacific, February 28, 2020; revised version July 26, 2023}

\begin{abstract}
The field of computer vision has greatly matured in the past decade, and many of the methods and techniques can be useful for astronomical applications. One example is in searching large imaging surveys for objects of interest, especially when it is difficult to specify the characteristics of the objects being searched for. We have developed a method using contour finding and convolution neural networks (CNNs) to search for Infrared Dark Clouds (IRDCs) in the \Sp\ Galactic plane survey data. IRDCs can vary in size, shape, orientation, and optical depth, and are often located near regions with complex emission from molecular clouds and star formation, which can make the IRDCs difficult to reliably identify. False positives can occur in regions where emission is absent, rather than from a foreground IRDC. The contour finding algorithm we implemented found most closed figures in the mosaic and we developed rules to filter out some of the false positive before allowing the CNNs to analyze them. The method was applied to the \Sp\ data in the Galactic plane surveys, and we have constructed a catalog of IRDCs which includes additional parts of the Galactic plane that were not included in earlier surveys.

\end{abstract}


\section{INTRODUCTION} \label{sec:intro}
\subsection{Infrared Dark Clouds}
Infrared dark clouds (IRDCs) are dense, cold clouds that appear as absorption features against the diffuse emission from the Galactic plane. Large numbers were first identified in observations from space-based instruments such as ISOCAM on ISO \citep{1996A&A...315L.165P,2001A&A...365..598H} and in the {\it Midcourse Space Experiment} (MSX) Galactic plane survey \citep{1998ApJ...508..721C,1999ESASP.427..671E} which had the sensitivity and spatial resolution to detect these compact, dark features in the mid-infrared. Studies of IRDCs showed them to be frequently associated with sites of massive star formation, and to contain cores and young stellar objects \citep[e.g.,][]{2005ApJ...630L.181R,2008ApJ...689.1141R,2006ApJ...653.1325S,2007ApJ...662.1082R,2006ApJ...641..389R,2006ApJ...653.1325S,2010ApJ...715..310R}. IRDCs have been identified as being the possible birthplaces of massive stars and stellar clusters \citep[see reviews by][and references therein]{2007ARA&A..45..339B,2018ARA&A..56...41M}, and so it is important to study the formation and evolution of these objects to understand this crucial phase in the star formation process.

\begin{figure}[ht]

\centering
	\includegraphics[height=0.4\textwidth]{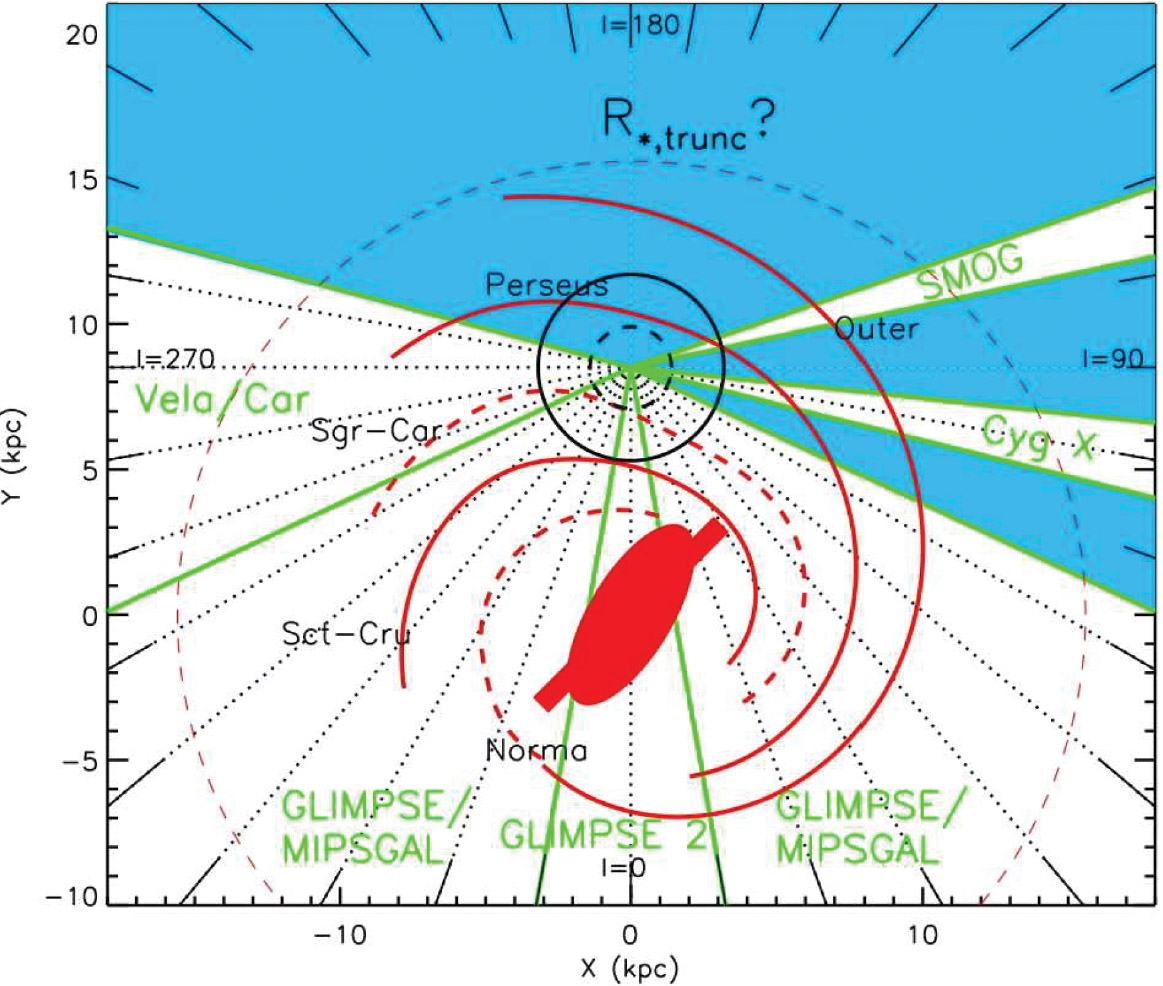}
	\caption{From \cite{2014Meade}, a schematic of the Galactic Plane showing the areas covered by the GLIMPSE surveys, with the GLIMPSE 360 survey in blue and others in white. The GLIMPSE 3D surveys cover regions extending in and out of the plane of the Galaxy in the lower part of this figure. The approximate positions of Galactic spiral arms \citep{1993ApJ...411..674T} are indicated in red. The radius marking the expected truncation or break in the exponential Galactic stellar disk is also shown with a dashed line. 	 
}\label{GPplot}
\end{figure}

A significant advance came with the \Sp\ Space Telescope \citep{Werner2004}, which provided a factor of $\sim$10 better resolution and a factor of $\sim$1000 better sensitivity than previous surveys at 8~\micron. The GLIMPSE survey \citep{2003PASP..115..953B,2009PASP..121..213C}, one of the \Sp\ Legacy surveys\footnote{\url{https://irsa.ipac.caltech.edu/data/SPITZER/docs/spitzermission/observingprograms/legacy/history/}} conducted early in the mission, mapped the $|l|<65$\degr, $|b| \le 1$\degr\ region of the Galactic plane with the IRAC instrument \citep{Fazio2004} at 3.6, 4.5, 5.8 and 8~\micron. The MIPSGAL survey \citep{2009PASP..121...76C} mapped a similar region with the MIPS instrument \citep{2004ApJS..154...25R} at 24 and 70~\micron. Based on these surveys, \citet[][hereafter PF09]{peretto09} conducted a search for IRDCs in the \Sp\ data using primarily 8~\micron\ opacity maps that they constructed from the available mosaics. They produced a catalog of over 11,000 clouds and up to 50,000 fragments within the clouds, indicating possible dense regions which could be forming cores. They also found many 24~\micron\ sources associated with the IRDCs, indicating active star formation. The improved resolution and sensitivity of \Sp\ allowed them to identify many clouds that were missed in the earlier MSX survey.

Later in the \Sp\ mission, further surveys by the GLIMPSE team and other groups filled in the full 360\degr of the Galactic plane (see Figure \ref{GPplot} and \S \ref{obs}). We were interested in searching these parts of the plane for IRDCs. However, a large part of the survey was conducted in the GLIMPSE 360 program \citep{2008sptz.prop60020W,2009AAS...21421001W} during the \Sp\ Warm Mission \citep{2010SPIE.7737E..1WM}, and so only IRAC 3.6 and 4.5~\micron\ data are available for those regions. Therefore our search technique would have to be different than those that used the 8~\micron\ IRAC band. We wanted to use an automated technique to search the large amount of data, for reasons of reproduceability and reliability. However, IRDCs can vary in size, shape, orientation, local background levels, and optical depth, and are often located near regions with complex emission from molecular clouds and star formation, which can make the IRDCs difficult to identify. False positives can occur in regions where emission is absent, rather than absorption from a foreground IRDC. We wanted a method that could maximize the identifications of true IRDCs and minimize false positives to produce a catalog of IRDCs for the entire Galactic plane.

\subsection{Contour Finding and Convolution Neural Networks}
\begin{figure*}[ht]
\begin{center} 
\includegraphics[width=0.8\linewidth]{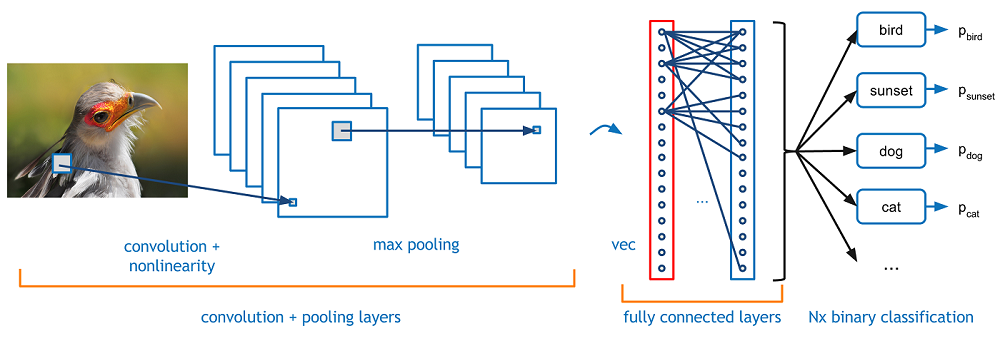}
\caption{From \citet{deshpande2016}, 
the diagram illustrates a convolutional neural network in its simplest form, where there are a predefined number of kernels with a fixed size that convolve over an image. The resulting image/matrix is then processed to reduce its size, in this illustration max pooling is used. The resulting matrix is then flattened and processed by a traditional neural network, which outputs the probability for each of the potential labels.}\label{CNN}
\end{center}
\end{figure*}

Convolutional neural networks (CNNs) are specialized deep neural networks that are widely used in image recognition and classifications. See for example \citet[][chapter 9]{Goodfellow-et-al-2016} for a general description of CNNs and their use in image analysis. They are better at capturing the features in an image through the application of relevant kernels, which convolve over the entire image. See Figure \ref{CNN} for a diagram of the CNN process. The kernels are fixed in size and numbers, but part of training a CNN is to optimize the values within the kernels. After each convolutional layer, the resulting matrices are usually downsized to simplify and speed up the computation.  This process is repeated a few times to optimize the values. Finally the matrices are flattened, converted into a one-dimensional array, and passed through a neural network to make the final judgement. Usually, and in our case, the output is a vector of probabilities, one value for a class. Because we want to use the CNN to decide whether the image contains an IRDC or not, we have the CNN output 2 values, one for the probability that the image is an IRDC and the other for not containing an IRDC. A CNN is trained in a supervised manner, meaning that it is fed a dataset of images with their corresponding labels, and the CNN is initialized with a random values for the kernels, weights and biases. The CNN then goes through the training set and calculates the loss at each step and adjusts all of the parameters and repeats the process until the error converges to a minimum value.

\begin{figure}[ht]
\centering
	\includegraphics[width=0.99\linewidth]{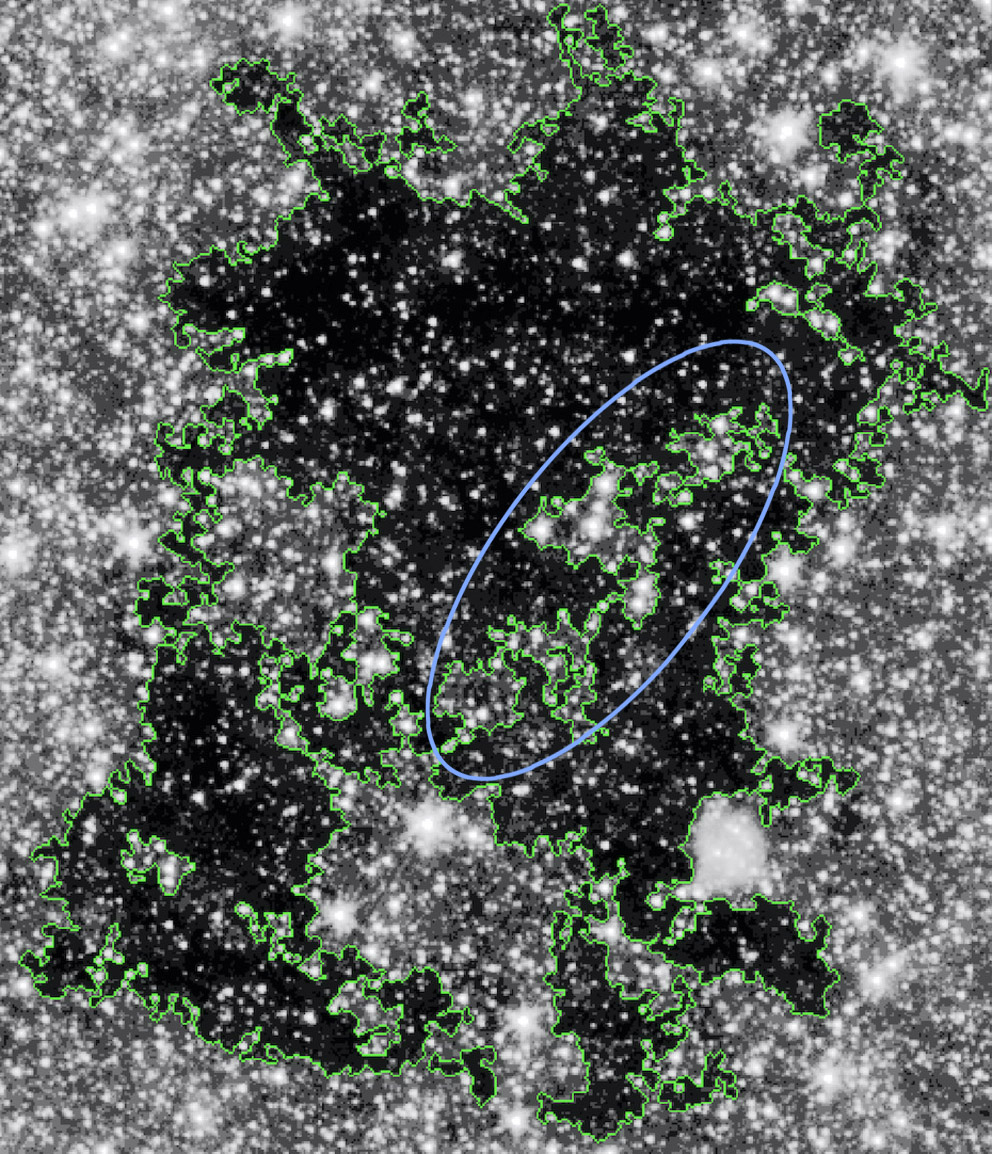}
	\caption{The IRDC G010.71--00.167 is outlined by the contouring finding method. However, within the blue oval there are contours found within the IRDC which in this and almost all cases contain clusters of bright stars. We are able to keep these contours from being identified as separate IRDCs through various types of filters as described in the text.}\label{contourExplain}
\end{figure}

We chose to use the MobileNet \citep{HowardZCKWWAA17} CNN model because it is computationally fast and can process our large images in a small amount of time.
 CNNs have been used before in astronomical applications; for example, they have analyzed images to automatically classify objects as stars or galaxies \citep{10.1093/mnras/stw2672}. Furthermore, they have been widely used outside of the astronomical community. For example, they have been employed to automatically classify animals in wildlife cameras \citep{NorouzzadehE5716}. 
However, there is a downside to CNNs; they cannot determine how many of a certain object exists or their location within an image. This is why we also employ a contour finding algorithm to find all closed figures, which is useful when searching for IRDCs because they are closed and distinct from the background. The contour finding method we used is called Topological Structural Analysis of Digitized Binary Images by Border Following \citep{SuzukiA85}. It takes in a binary image, and by following the edges, it finds closed loops. It finds an edge and then checks its neighboring pixels to see if any of them are edges, and it keeps moving around the border in this manner. If it has reached the starting point, then it identifies it as a closed loop, or a contour. This is a popular method, which is why its incorporated into OpenCV, however it can also find closed loops within other closed loops. Unfortunately, this aspect of finding contours within contours was not useful for us, because almost all of the sub contours ended up being bright regions within the IRDCs caused by star clusters. This is shown in Figure \ref{contourExplain}. The contour method locates IRDCs and a sizeable amount of non-IRDC objects. This is why we use a series of filters and finally a CNN to decipher which of the contours are IRDCs.

\section{Observations}\label{obs}
\begin{figure*}[ht]
\begin{center}
  \includegraphics[width=5.5in]{{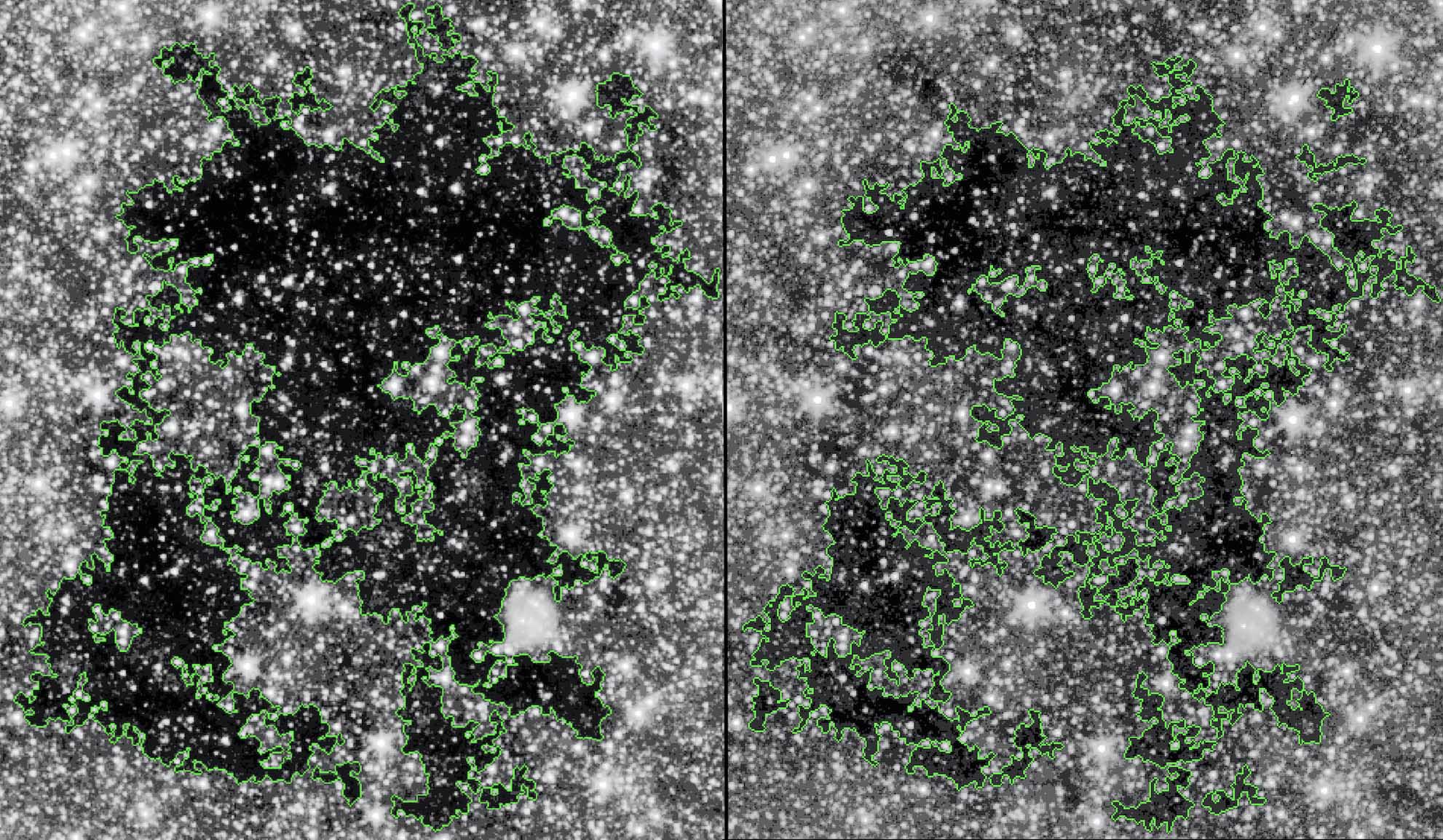}}
  \caption{The IRDC G010.71--00.167 is shown in the 3.6 \micron\ image on the left and the 4.5 \micron\ image on the right. The images were processed with histogram equalization and contour finding using the same parameters. The green outline is the output of the contour finding method. It is clear that in the 3.6 \micron\ image the contours better represent the full size of the IRDC, and the IRDC is not broken up into as many separate contour areas as in the 4.5 \micron\ image.} \label{Channel}
\end{center}
\end{figure*}

The mosaics used here are based on images obtained with the IRAC instrument on \Sp.
We used the mosaic images created by the GLIMPSE team for the GLIMPSE and other surveys conducted during the \Sp\ mission. These include the original GLIMPSE survey \citep[GLIMPSE I -- ][]{2003PASP..115..953B,2009PASP..121..213C}, GLIMPSE II  \citep{2005sptz.prop20201C}, GLIMPSE 360 \citep{2008sptz.prop60020W,2009AAS...21421001W}, GLIMPSE 3D \citep{2006sptz.prop30570B,2007AAS...211.1409B}, Deep GLIMPSE \citep{2011sptz.prop80074W}, Vela-Carina \citep{2007sptz.prop40791M,2009ApJ...707..510Z}, SMOG \citep{2008sptz.prop50398C}, and Cygnus-X \citep{2007sptz.prop40184H,2009AAS...21335601H}. The location of each of the surveys as a function of Galactic longitude is shown in Figure \ref{GPplot}. The 1\farcs{2}/pixel images in Flexible Image Transport System (FITS) format were used, and the background-matched and gradient-corrected or latest versions available were downloaded from the IRSA database\footnote{\url{https://irsa.ipac.caltech.edu/data/SPITZER/GLIMPSE/\\ overview.html}}. 
The GLIMPSE I, GLIMPSE II, GLIMPSE 3D, Vela-Carina, SMOG, and Cygnus-X surveys were conducted in the four IRAC bands (3.6, 4.5, 5.8, and 8~\micron; Channels 1--4, respectively) during the cryogenic mission. The GLIMPSE 360 and Deep GLIMPSE surveys, which included most of the new area that we wished to survey for IRDCs, were conducted during the \Sp\ Warm Mission and data were obtained only in the 3.6 and 4.5~\micron\ bands. We therefore conducted tests to determine whether to use Channel 1 or 2, or some combination, for our IRDC search. After some tests and examining the mosaics, we chose to use Channel 1 images only because the IRDCs appear more visually prominent compared to their appearance in Channel 2, possibly due to the higher dust extinction at 3.6~\micron\  \citep[e.g.,][]{2007ApJ...663.1069F} and the brighter stellar background. We looked at the same regions in both channels, with the images processed by histogram equalization and contour finding. The results were consistently better (fewer missed objects, more IRDC area identified) in Channel 1 than 2 because the IRDCs are more prominent in the Channel 1 images and the contour finding algorithm was able to find and more fully map the spatial extent of the IRDCs. A representative example is shown in Figure \ref{Channel}. The IRDC in the Channel 2 image is broken up into more contours than the same object in Channel 1. Furthermore, the contour finding procedure captured more of the size and area of the IRDC in the Channel 1 image.

\section{Search Method}

\subsection{Overview of Procedure}
We implemented our procedure in Python3, and have uploaded the code to a GitHub repository\footnote{\url{https://github.com/jyopari/IRDC}}. Our code makes use of the \verb|Astropy(3.1.2)|\footnote{\url{https://www.astropy.org/}}, \verb|Pillow(5.2.0)|\footnote{\url{https://pillow.readthedocs.io/en/5.2.x/index.html}}, \verb|OpenCV(3.4.2.17)|\footnote{\url{https://opencv.org/}}, and \verb|Tensorflow(2.0.0a0)|\footnote{\url{https://www.tensorflow.org/}} libraries. The first steps of our procedure were to apply image filtering techniques and contouring finding in order to locate all the closed figures in the mosaic. We obtained the bounding boxes of the figures and evaluated which ones should be kept based on their characteristics. Finally, the remaining bounding boxes were then sent to a classifier which provided the final classification and determined whether or not each figure is an IRDC. We describe these steps in more detail below.

\subsection{Processing Steps}\label{parmchoice}
\begin{figure}[h]
  \includegraphics[width=\linewidth]{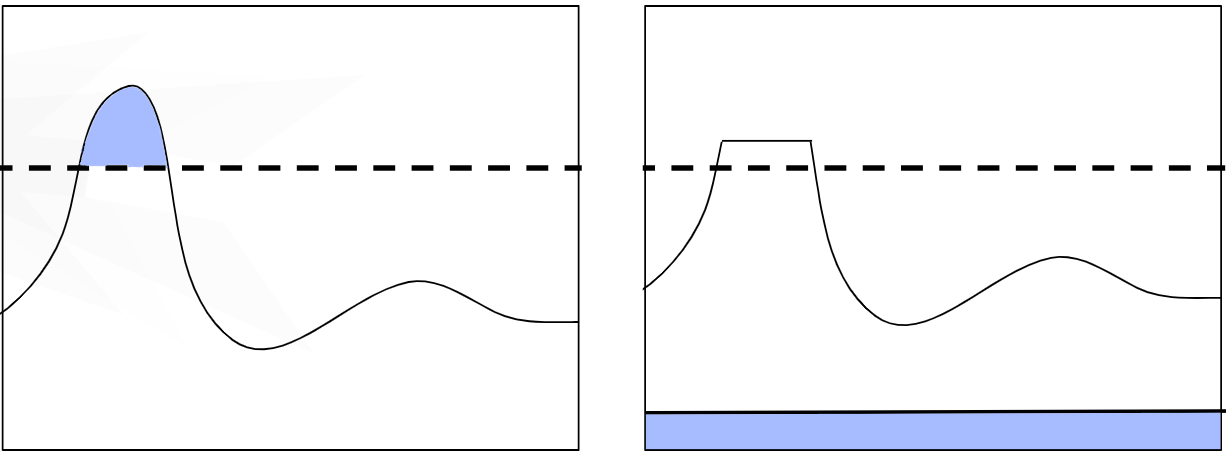}
  \caption{The histogram on the left and right represent the color frequency  for a mosaic before and after the clip limit transformation. The dashed horizontal line that passes through one of the peaks in the histogram on the left represents the clip limit value. If there are values that exceed the clip limit value (shown in blue in the left panel), then they are evenly added to the all the bins (the blue region in the right figure), resulting in a constant increase to all the bins after the transformation.} \label{clipLimit}
\end{figure}

\begin{figure}[ht]
  \scalebox{1}[1]{\includegraphics[width=\linewidth]{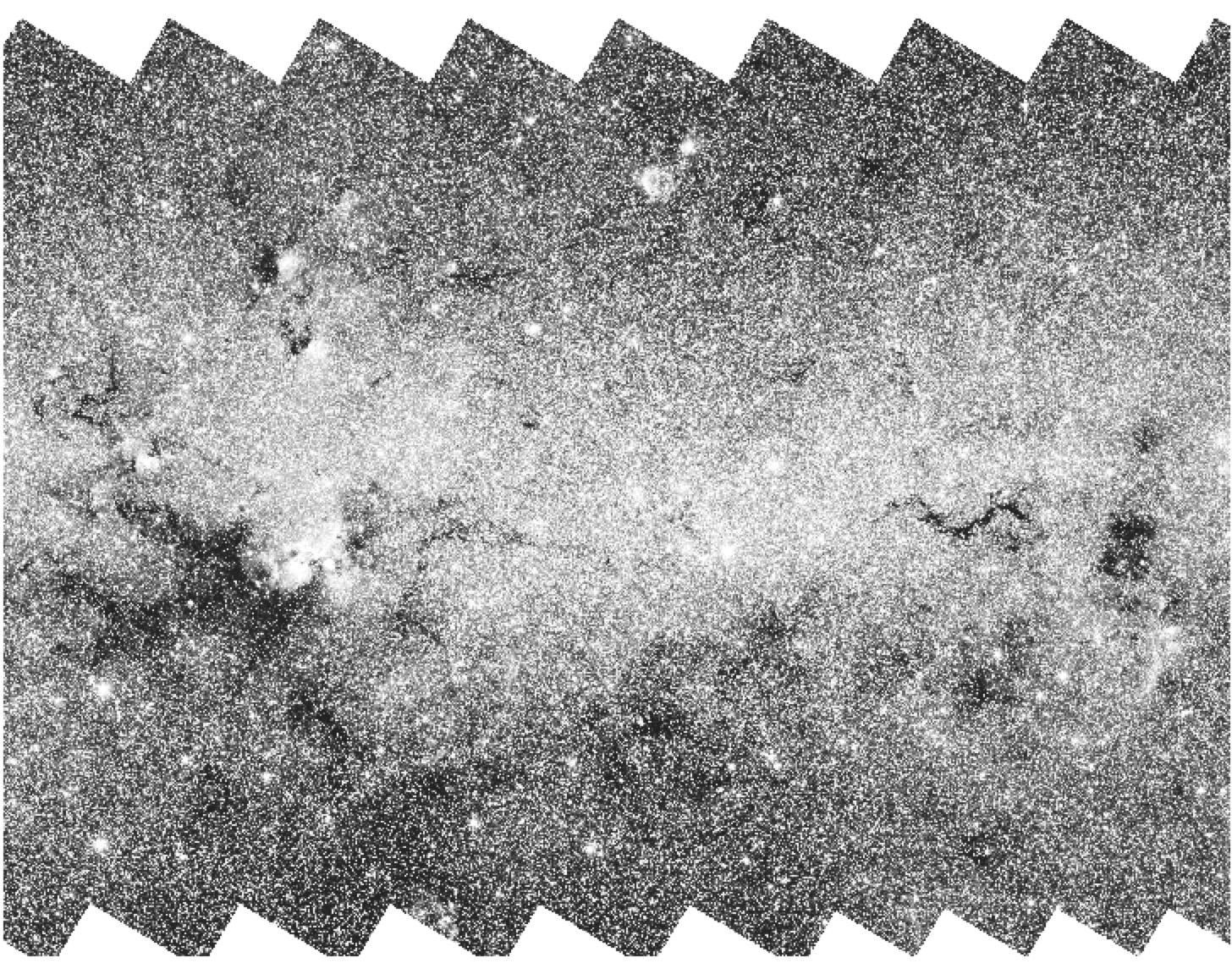}}
  \includegraphics[width=\linewidth]{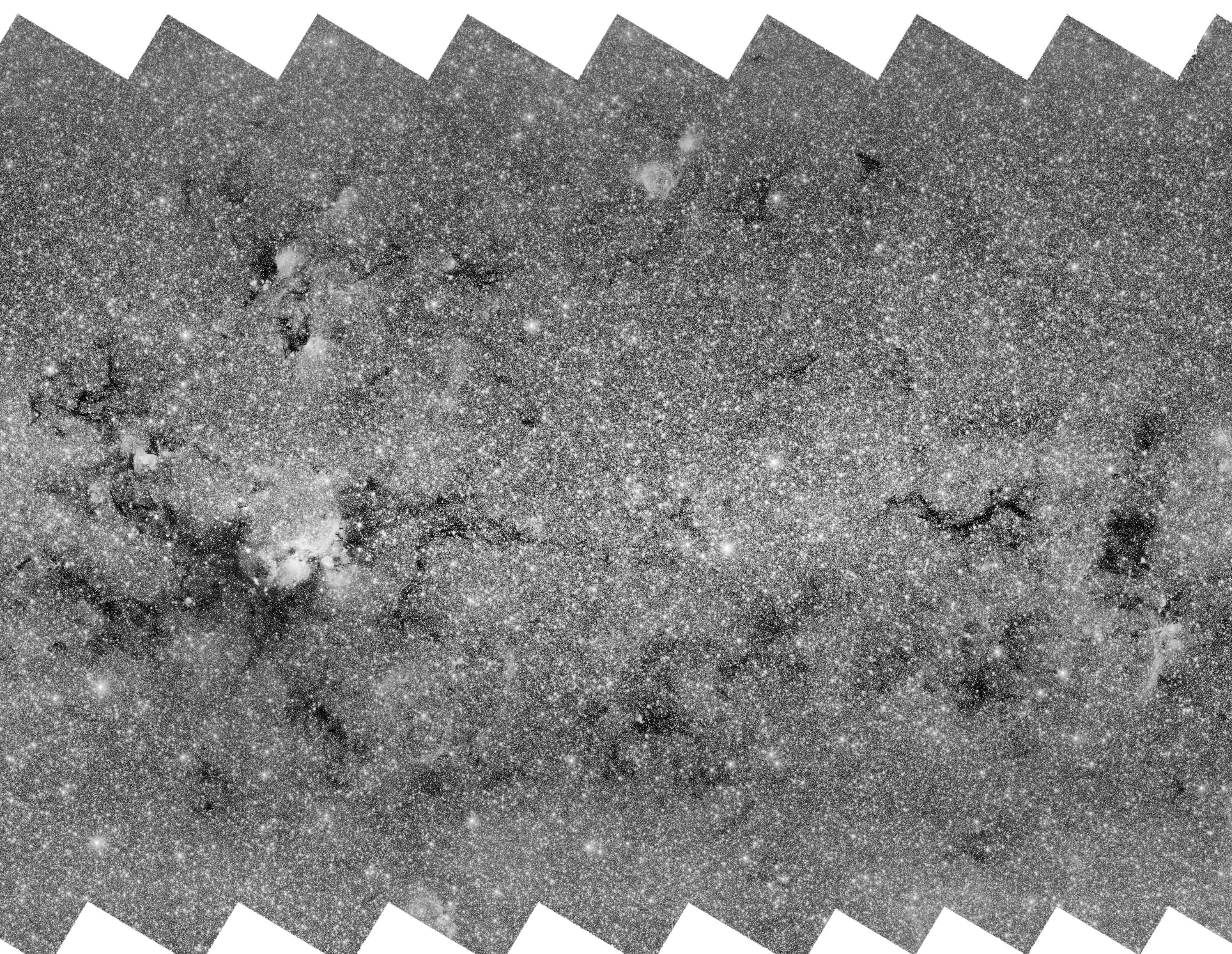}
  \caption{The mosaic GLM\_01200+0000\_I1 shown on the top was processed using histogram equalization. The bottom image is the top image after the contrast limited adaptive histogram equalization technique was applied. The adaptive histogram technique helps reduce the effects of the large variations in background level across the image, and clarifies the structure of the image near the top and bottom edges where the background is lower than the center.} \label{histogram}
\end{figure}

For each mosaic in the dataset, we read the FITS-format file into an image array of intensity values using the \verb|Astropy| library. The values were then scaled by a multiplicative factor so that when the flux density values (in MJy sr$^{-1}$) are truncated and written to a file with values in the range 0 -- 255, there will be enough dynamic range in the lower flux levels to detect the IRDCs. We used a factor of 5 or 10, depending on the background level in the particular dataset being processed. The values used are given in Table \ref{params}. We then used the \verb|Pillow| functions \verb|fromarray()| and \verb|convert()| to convert the data into a grayscale image and saved the image in PNG format. The image was then processed using the Contrast Limited Adaptive Histogram Equalization (CLAHE) technique \citep{Zuiderveld:1994:CLA:180895.180940} through OpenCV's \verb|createCLAHE()| function. Regular histogram equalization looks at the distribution of all the pixel values in an image and stretches it over the entire domain of values. This will flatten any high density regions in the distribution, thus increasing the contrast in the image. However, one drawback of using histogram equalization is that it transforms the histogram of the whole image. This will not provide as much clarity in an image where different regions within the image are distinct. In our case, there is an especially strong background gradient across the Galactic plane in mosaics from GLIMPSE I and II, and many bright regions that can dominate the high end of the intensity histogram. Because of this, we employed the CLAHE technique\footnote{\url{https://docs.opencv.org/master/d5/daf/tutorial_py_histogram_equalization.html}}. This method uses histogram equalization on small subsets of the image, so that parts of the image that are vastly different in character do not influence the local equalization process. We used a tile size of 8 by 8, which means that histogram equalization is performed on a square with a length of 8 pixels. A key component of the CLAHE process is the clip limit. If there is a region in an image where the values are very similar, there it will likely produce a peak as shown on the left side histogram in Figure \ref{clipLimit}. Histogram equalization tends to excessively spread the peak, to make it more uniform, which will create an unwanted high contrast in that region. To solve this, before histogram equalization is applied, bins in the histogram that are larger than a threshold, called the clip limit, will be reduced to stay at the threshold level. However, since reducing the size of some bins will cause the total sum of the bins to decrease, the amount that is taken away from the bins above the clip limit is then evenly distributed as shown in Figure \ref{clipLimit}. This step was performed in order to minimize the effects of the changing background levels in different parts of the image, as seen in the top image of Figure \ref{histogram}. The greater contrast that CLAHE produces compared to regular histogram equalization is illustrated in the lower image of Figure \ref{histogram}. 

After CLAHE was applied, we used Gaussian blurring to smooth the image and reduce noise before a binary threshold was applied; both processes were conducted via \verb|OpenCV's| functions \verb|GaussianBlur()| and \verb|threshold()| respectively. The Gaussian kernel's dimensions were 5$\times$5. Using those dimensions, OpenCV calculated $\sigma$ using the following expression: $ \sigma = 0.3*(\frac{size-1}{2}-1) + 0.8 $. Our kernel size was 5 by 5 pixels, so $\sigma = 1.1$. Below is a rounded version of the kernel:
\par
\begin{center}
$
\begin{bmatrix}
$.01	.02	.03	.02	.01$\\
$.02	.06	.09	.06	.02$\\
$.03	.09	.14	.09 .03$\\
$.02	.06	.09	.06	.02$\\
$.01	.02	.03	.02	.01$\\
\end{bmatrix}
$
\end{center}

After the image is smoothed, it is then converted to a binary image before sending it to the contour finding algorithm. We experimented with various constant values for the threshold cutoff, and we found that values around 50 provided the most reliable results across the mosaics from the different surveys. The threshold can be related to the original image intensity through the scaling factor used for the region, given in Table \ref{params}. 
We then employed Topological Structural Analysis of Digitized Binary Images by Border Following \citep{SuzukiA85} via OpenCV's \verb|findCountours()| function. This method uses border following in order to locate closed figures in a binary image. 
\par
\begin{figure}[ht]
  \includegraphics[width=\linewidth]{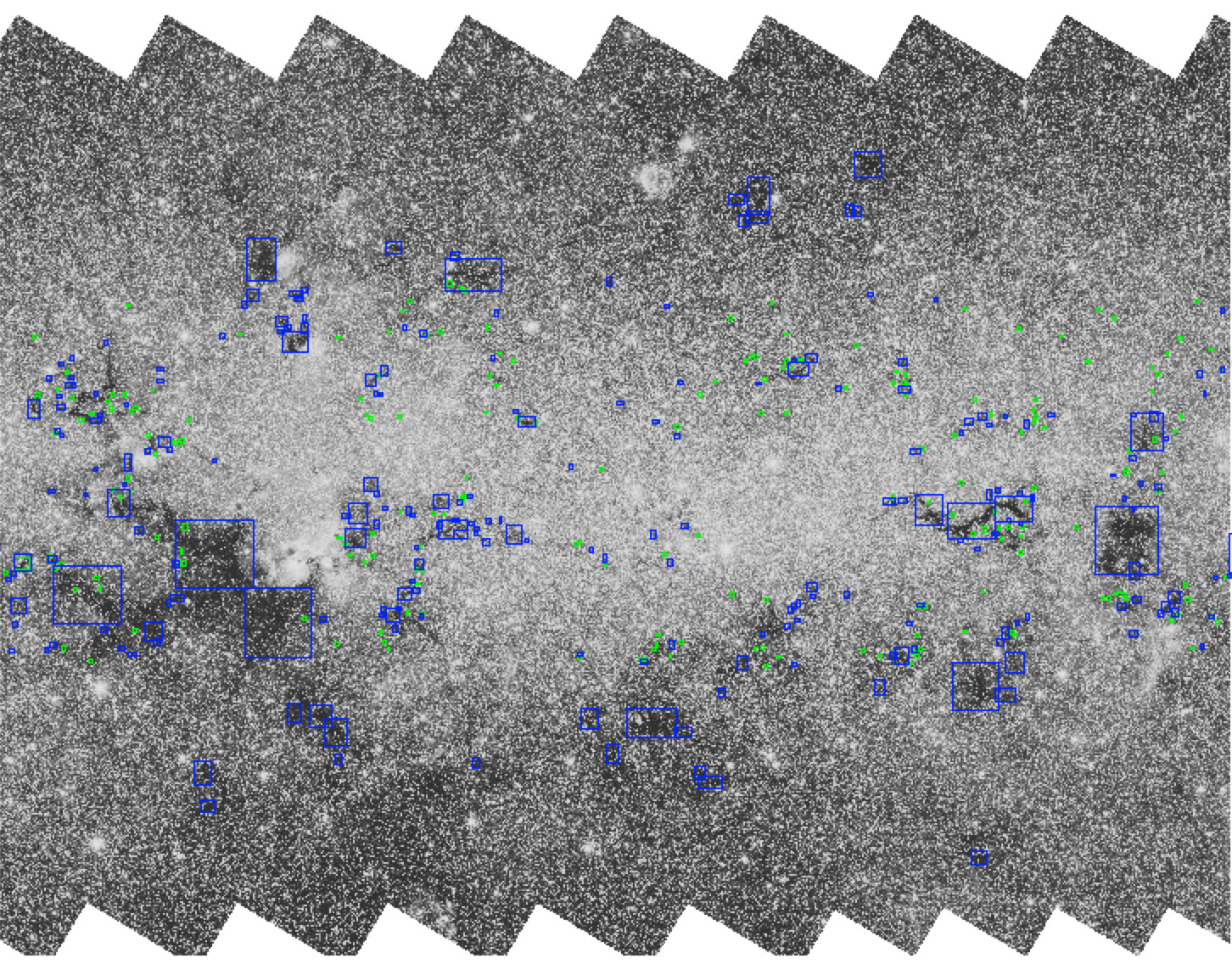}
  \caption{The mosaic GLM\_01200+0000\_I1 showing the adverse effects when increasing the number of contours for processing from 1000 to 2000. The blue boxes represents the IRDCs found when 1000 contours are evaluated, whereas the green boxes represent the additional IRDCs found when 2000 contours are processed. The green areas are typically very small regions that do not correspond to real IRDCs in the image, or areas in or near larger IRDCs already identified.}
  \label{threshold}
\end{figure}
\begin{figure}[h]
  \includegraphics[width=\linewidth]{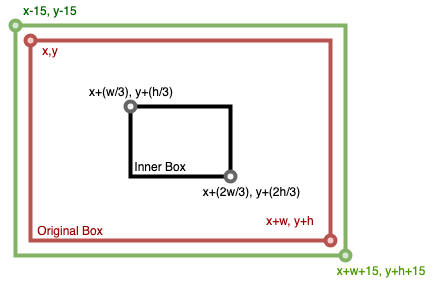}
  \caption{A graphical representation of the process to determine the difference between the IRDC core and surrounding region. The red box represents the original bounding box for the contour, whereas the green box represents the expanded box used for determining the local level. The black box shows the region used for a measure of the IRDC depth. The difference value is calculated by subtracting the outbox's average from the inner box's average.} \label{_difference}
\end{figure}

We then obtained the bounding box for each contour using the \verb|boundingRect()| function from OpenCV. The bounding boxes were then sorted based on their area, and we only kept the first 1000 largest bounding boxes in each mosaic. We found that values exceeding 1000 produced exceedingly more false positives than true positives, as illustrated in an example image in Figure \ref{threshold}. We then established a criterion to evaluate a potential IRDC based on the intensity difference between the inner part of the IRDC and its immediate surroundings. Because IRDCs are usually darkest near the center, we shrunk the original bounding box dimensions by a third while not moving its center and  calculated the average intensity within the bounding box. Then we defined another box which is 30 pixels larger in both dimensions with the same center, and calculated its average intensity and subtracted the average intensity of the smaller box from the larger one. The definition of inner and outer boxes is shown in Figure \ref{_difference}. We then created three subsets, each with their respective filter. The first subset consisted of bounding boxes whose contour areas were greater than 30,000 pixels, the second subset received bounding boxes whose contour areas were between 30,000 and 3000 pixels, inclusive, and the third subset was defined as having bounding boxes whose contour area was less than 3000 pixels. The contour area was calculated through OpenCV's \verb|contourArea|. We defined the parameters $T_{0}$, $T_{1}$, $T_{2}$ for the large, medium, and small subsets, respectively. The parameter value is the threshold for the intensity difference between the larger and smaller boxes as defined above, meaning for example if a bounding box has a contour area of 40,000 pixels, then it would need an intensity difference greater than $T_{0}$ for it to not be eliminated from consideration. 

These cutoffs were applied to the list of bounding boxes, and the remaining boxes were then processed by a \verb|MobileNets| \citep{HowardZCKWWAA17}. We used MobileNets, implemented in \verb|Tensorflow|, for their proven accuracy and smaller architecture, which increased the computational speed. The MobileNet was trained on IRDCs that had bounding boxes smaller than 30,000 pixels. We used objects from the PF09 catalog as a guide for the GLIMPSE I and II regions for our training data. We visually confirmed that the PF09 objects selected appeared to be true IRDCs. Our training set for each classifier comprised of approximately 50 bounding boxes containing dark clouds and 50 bounding boxes that contained no IRDCs. We did not need a large training set because the MobileNets came pretrained on a general image dataset called ImageNet \citep{imagenet_cvpr09}. This means that they have already been trained from scratch on a large dataset, and they have ``learned" which features to observe in an image in order to classify it. We applied transfer learning, meaning that we trained the trained model for our dataset, i.e., the final layer of the network\footnote{\url{https://www.tensorflow.org/hub/tutorials/image\_retraining}}. 

The  classifier was trained on the smaller IRDCs, and we used different thresholds on its output confidence depending on the size of the IRDC. Specifically, we used 30000 pixels as the threshold to determine if an IRDC should be considered small or not. We did not use contour area to determine which classifier viewed which bounding box, instead we estimated the area by multiplying the width and height of the box. Each classifier returned the probability of the image inside the bounding box containing an IRDC. The classifier returned the probability of the image inside the bounding box containing an IRDC. We defined two parameters $C_{0}$, $C_{1}$, which are probability thresholds for small and large candidate IRDCs respectively that the returned probability needs to exceed in order for the bounding box to be recorded into the catalog. Once a bounding box passed all the conditions in the pipeline, its coordinates, width, and height were appended to a region file in Galactic coordinates. We used the \verb|all_pix2world()| function from Astropy to obtain the Galactic coordinates from the pixel coordinates of the image. We only needed one classifier since it showed reliable performance on candidate IRDCs of various sizes.  
\par
We also calculated an estimate of the maximum absorption depth of the IRDC. We created a rectangle 30 pixels wider and taller that shared the same center as the original bounding box. We then calculated the median intensity value of the ''border'' region (the 15 pixel wide region surrounding the original box), and subtracted from that the minimum non-negative intensity of a pixel within the original bounding box. We then divided the difference by the median border value as a measure of the maximum depth of the feature, with larger values indicating deeper absorption.

\subsection{Parameter Optimization}
\begin{figure}[h]
  \includegraphics[width=\linewidth]{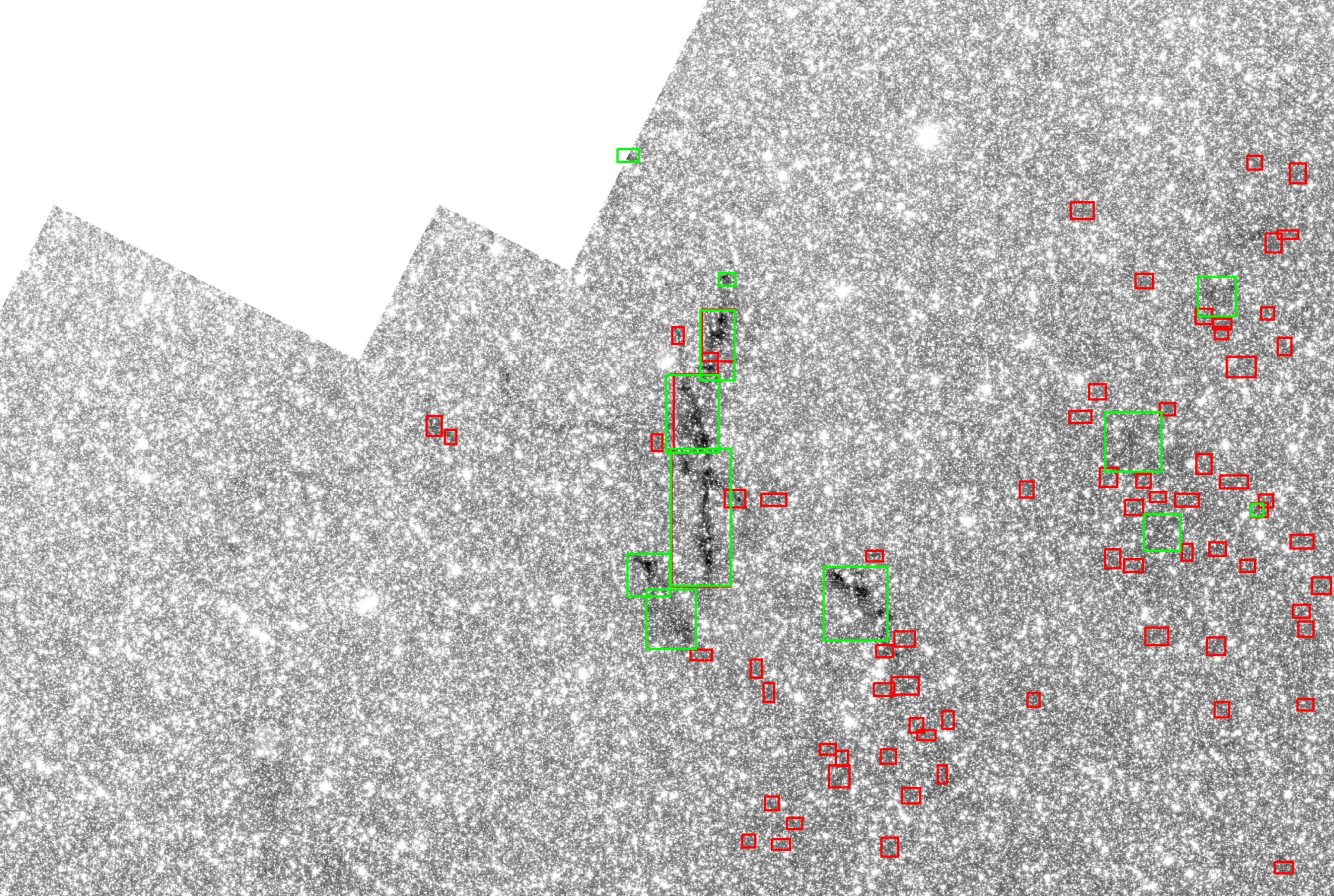}
  \caption{A section of the mosaic from the 3D region, 00000-0300 is shown with objects identified using a value for $T_{2}$ of 40 in red and $T_{2}$ = 85 in green. A total of 457 and 16 objects were found in this mosaic for values of 40 and 85, respectively. The lower $T_{2}$ value resulted in many more small objects being identified which are not likely real IRDCs.}
  \label{T_Change}
\end{figure}

The algorithm parameters $T_{0}$, $T_{1}$, $T_{2}$, $C_{0}$, and $C_{1}$ needed to be optimized to reliably locate IRDCs in the mosaics. We found that one set of values would not work for the entire dataset, since the data were taken with different exposure times and covered regions with large differences in background levels and complexity. Thus, for each of the surveys listed in \S \ref{obs}, we systematically varied the parameters and evaluated the results for a small subset of the mosaics in the survey to optimize the parameter values. After observing the difference of the bounding boxes from a set of mosaics, we determined a base starting point of $T_{0}=6, T_{1}=15, T_{2}=40$. From there we then let $C_{0},C_{1} \in [0, 0.25, 0.5, 0.75]$. Since these parameters impact results independently, when finding the optimum values for $C_{0}$ and $C_{1}$, we turned one off and ran the classifier using all the values. Once we found the optimum value for one parameter, we then turned it off and repeated the process for the other parameter. We then repeated this process for each mosaic in the sample set and obtained a set of $C_{0}$ and $C_{1}$. We averaged them, and the result was the parameter for the whole dataset for that survey. However, in some regions the results were not improved when changing the values of $C_{0}$ and $C_{1}$. We then experimented with changing $T$ until we found optimum values for $T_{0}$, $T_{1}$, $T_{2}$. An example of the effect of changing these parameters is shown in Figure \ref{T_Change}. The final parameters we used for each survey are given in Table \ref{params}.

\begin{deluxetable}{ccccccc}
\caption{Parameters Used for Survey Areas}\label{params}
\tablehead{& & \multicolumn{2}{c}{Probability} & \multicolumn{3}{c}{Depth}\\
& \colhead{Scaling} & \multicolumn{2}{c}{Cutoff} & \multicolumn{3}{c}{Cutoff}\\
\colhead{Region} & \colhead{Factor} & \colhead{$C_{0}$} & \colhead{$C_{1}$} & \colhead{$T_{0}$} & \colhead{$T_{1}$} & \colhead{$T_{2}$}}
\startdata
 360     & 10 & 0.25 & 0.25 & 15 & 15 & 40  \\
 3D      & 10 & 0.5 & 0.0 & 6 & 6 & 85 \\
 CYGX    & 10 & 0.25 & 0.25 & 20 & 20 & 40  \\
 Deep    & 10 & 0.5 & 0.0 & 20 & 20 & 55  \\
 I       & 5 & 0.25 & 0.0 & 6 & 6 & 45\\
 II      & 5 & 0.5 & 0.0 & 6 & 6 & 40  \\
 SMOG    & 10 & 0.25 & 0.25 & 20 & 20 & 40  \\
 VelaCar & 10 & 0.25 & 0.25 & 25 & 25 & 55 \\
\enddata
\end{deluxetable}

\subsection{Further Processing}\label{method}
\begin{figure}[h]
 \includegraphics[width=\linewidth]{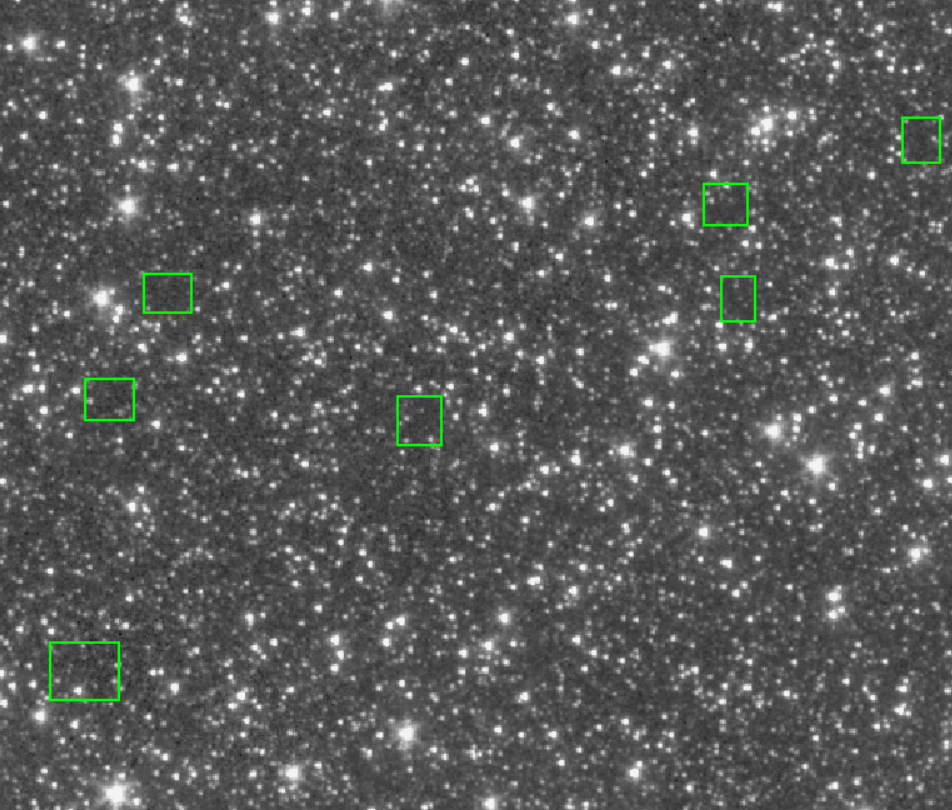}
  \caption{A sample section of the mosaic GLM\_04500+0000, showing several small regions initially identified in the processing but removed based on the small area cutoff level. The regions identified seem to be areas of low background between stars rather than due to absorption by an IRDC.}
  \label{smIRDCs}
\end{figure}

Once our initial catalog was generated, we applied several further processing steps to produce the final IRDC catalog. First, the contour algorithm had found contours within contours, which were not necessary for identifying the IRDCs, and led to duplicate identifications. In addition, because the mosaics in the survey overlap slightly, some IRDCs are identified in more than one image and produce duplicate identifications. As a result, we wrote another program which sifted through all the IRDCs in the catalog and removed any bounding boxes which had their centers within another bounding box, and their maximum extent did not exceed the larger bounding box by 10$''$ in any direction. In each case the smaller of the boxes were removed. We also applied a lower IRDC contour area cutoff of 300 pixels (432 arcsec$^2$) because it appeared that objects of this size or smaller were either false positives, or were identifying small fragments of IRDCs that were already in the catalog as larger IRDCs. This effect is illustrated in Figure \ref{smIRDCs}, which shows several small regions that were identified in the initial processing. The regions seem to be locating spaces between brighter stars, rather than true IRDCs. There are many similar regions in the image, and the identified areas do not stand out as distinctly different. This effect is a function of the IRAC instrument resolution and sensitivity, combined with the distribution of stars in the field.   By setting a lower limit to the size of the IRDCs in the catalog, we can minimize the inclusion of these regions that do not represent true IRDCs. Because of this effect, it is likely that a number of false positives remain in the catalog and the percentage is higher amongst the smallest IRDCs in the catalog.\\

\section{Results and Discussion}\label{results}
\begin{figure}[h]
  \includegraphics[width=\linewidth]{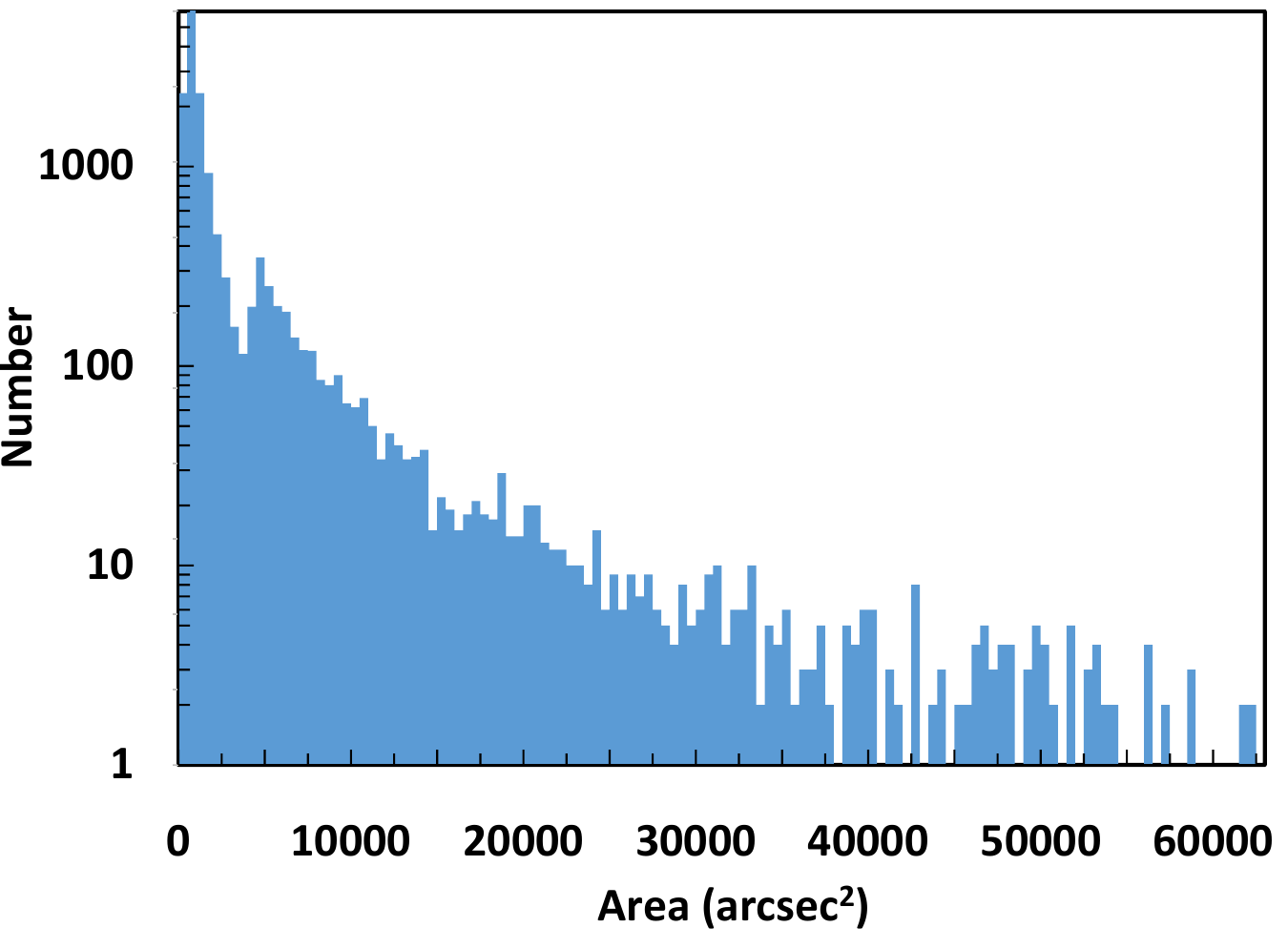}
  \caption{The distribution of IRDCs as a function of contour area. Over 84\% of the IRDCs have an area of 4000 arcsec$^2$ or less. The maximum area is 393,870 arcsec$^2$, and there are 106 IRDCs with areas higher than 60,000 arcsec$^2$.}  
  \label{Histogram}
\end{figure}

Using our procedure as described above, we find a total of 18,845 objects. We will refer here to the objects in our catalog as IRDCs, however they should be considered as candidate IRDCs -- not all of the objects are true IRDCs, and the catalog does not contain a complete list of all true IRDCs in the survey area (see \S \ref{limitations} for more discussion of the limitations of our method). Our catalog is shown in Table \ref{IRDCcat}, which gives the name, Galactic coordinates, size of the bounding box in arcseconds of Galactic longitude and latitude, depth ratio of the IRDC, probability value returned by the classifier, and the area within the contour of each IRDC in arcsec$^2$. The depth ratio is the difference between the median 3.6~\micron\ flux density around the outside of the IRDC box and the non-negative minimum level of a pixel within the box divided by the median outside of the box. Therefore larger numbers represent deeper absorption values. The final column in the table lists the number of objects from the PF09 catalog that have their centers within the borders of the IRDC box defined in the table, for those objects that fall within the region surveyed by PF09. A histogram of the distribution of IRDC sizes is shown in Figure \ref{Histogram}.

\begin{figure*}[ht]
  \includegraphics[width=\linewidth]{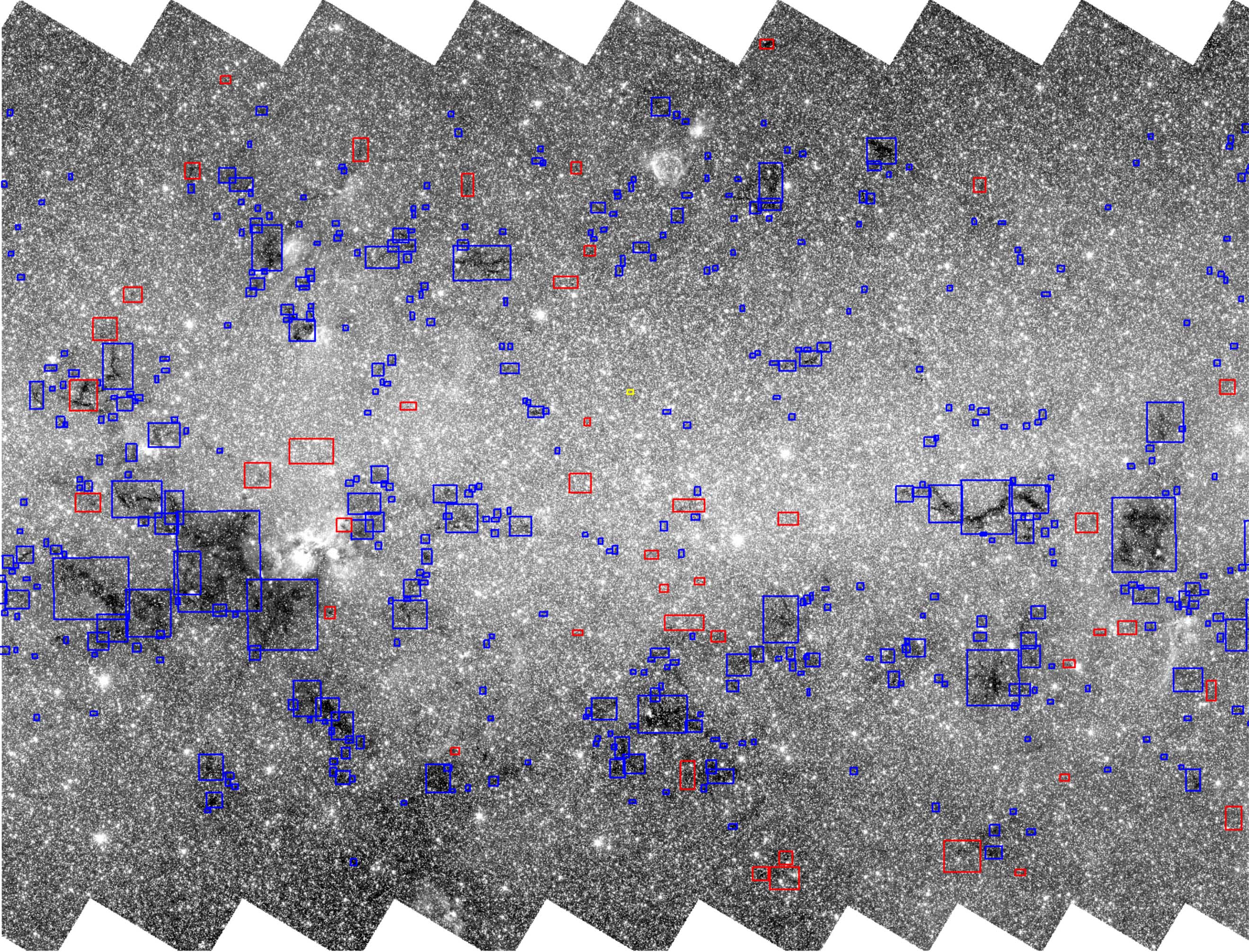}
  \caption{The mosaic GLM\_01200+0000 which was used as our test mosaic from the GLIMPSE I region. The blue boxes represents the true positives, where the red boxes represents the false negatives (IRDCs that were not identified by our process). The yellow boxes represents the false positives.}
  \label{Accuracy}
\end{figure*}

\begin{figure*}[ht]
  \includegraphics[width=\linewidth]{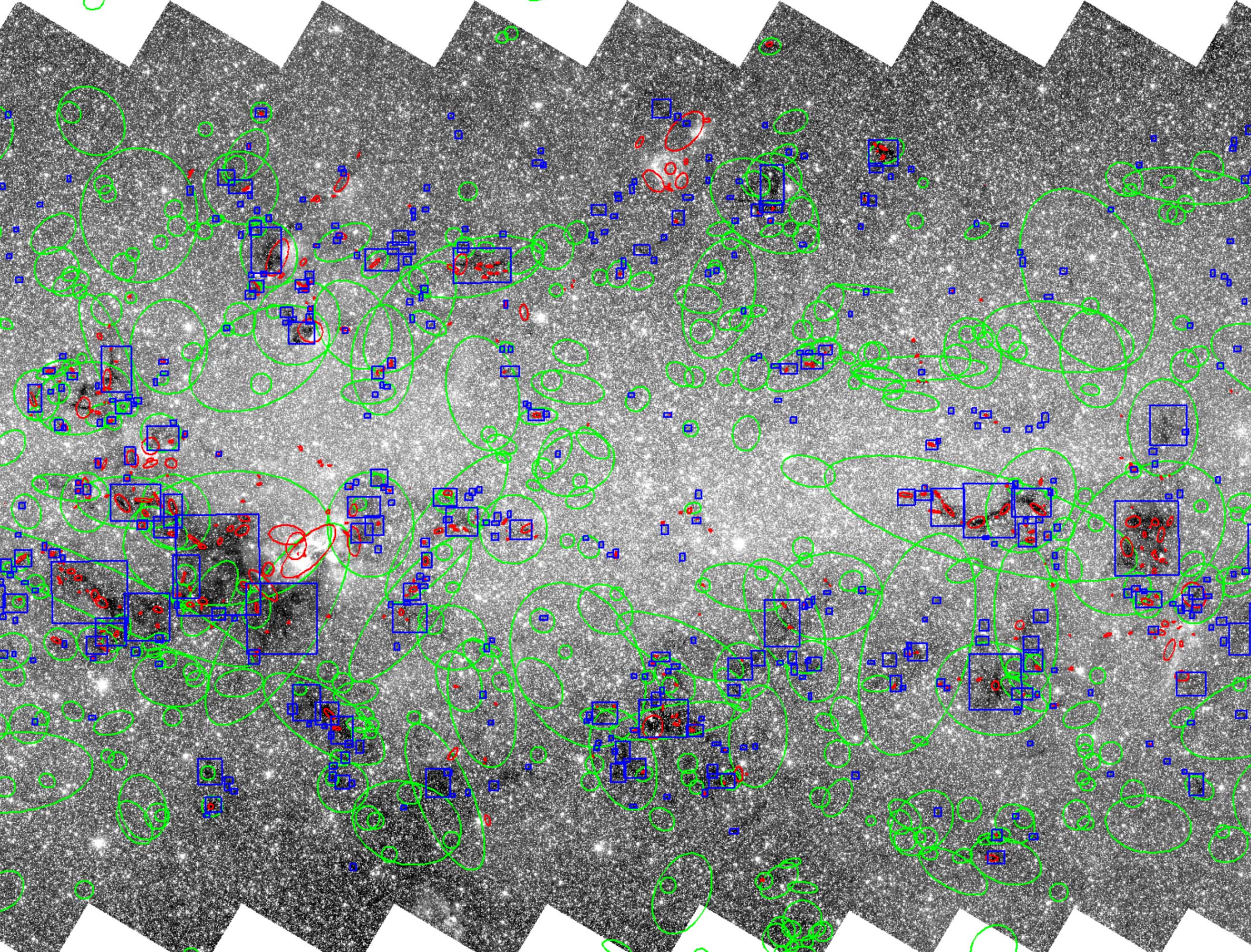}
  \caption{The GLM\_01200+0000\_I1 mosaic with blue rectangles for our catalog. We also show objects from the PF09 catalog in red and the MSX catalog \citep{Simon_2006} in green. The objects are shown as ellipses here, with the major and minor axes and position angle based on the IRDC sizes and angles given in those catalogs. }
  \label{Overlay}
\end{figure*}

We list the probability values in the table, although these did not seem to correlate well with the quality of the IRDC identification. For example, some of the large IRDCs identified in the field shown in Figure \ref{Accuracy} had probabilities in the range 0 -- 0.3, yet all seem to be definite IRDCs that were also identified in the prior surveys. Many other objects have a probability of 1.0, even some of the small IRDCs that are probably not as reliable. 

\begin{deluxetable*}{crrrrrrrr}
\tablecaption{IRDC Catalog\label{IRDCcat}}
\tablehead{& \multicolumn{2}{c}{Galactic Coordinates} &\multicolumn{2}{c}{IRDC Size}& \colhead{Depth}&\colhead{Proba-} &\colhead{Contour} & \colhead{PF09}\\
\cmidrule(lr){2-3}\cmidrule(lr){4-5}
\colhead{Name} & \colhead{Longitude} & \colhead{Latitude} & \colhead{(Longitude)} & \colhead{(Latitude)} & \colhead{Ratio\tablenotemark{a}} &\colhead{bility}& \colhead{Area}& \colhead{Objects\tablenotemark{b}}\\ 
\colhead{} & \colhead{(deg)} & \colhead{(deg)} & \colhead{(arcsec)} & \colhead{(arcsec)} & & & \colhead{(arcsec$^2$)} &  } 
\startdata
G000.01+00.275&	0.0095&	0.2752&	160.8&	138.0&	0.78&	1&	8428&	\nodata\\
G000.01$-$00.601&	0.0055&	-0.6012	&37.2	&62.4	&0.61	&1	&661&\nodata	\\
G000.02+00.359&	0.0152&	0.3592&	45.6&	44.4&	0.53&	1&	456&	\nodata\\
G000.02+02.797&	0.0249&	2.7972&	37.2&	42.0&	0.50&	1&	518&	\nodata\\
G000.02$-$00.433&	0.0195&	-0.4328&	142.8&	44.4&	0.92&	1&	2139&	\nodata\\
G000.03$-$00.349&	0.0342&	-0.3488&	48.0&	42.0&	0.71&	1&	596&	\nodata\\
G000.05+00.285&	0.0528&	0.2855&	73.2&	57.6&	0.59&	1&	1666&\nodata\\
G000.06+00.174&	0.0635&	0.1738&	52.8&	50.4&	0.77&	1&	1172&\nodata\\
G000.06$-$00.427&	0.0648&	-0.4272&	87.6&	45.6&	0.76&	1&	1607&\nodata\\
G000.07+00.215&	0.0735&	0.2152&	58.8&	40.8&	0.52&	1&	850&\nodata\\
\enddata
\tablenotetext{a}{The ratio of the difference between the median 3.6~\micron\ flux density around the outside of the IRDC box and the minimum level of a pixel within the box to the median outside of the box (see \S \ref{results}).}
\tablenotetext{b}{Number of objects from \cite[PF09;][]{peretto09} that have their centers within the IRDC box as defined in this catalog. An ellipsis is shown in this column for those objects that were not in the region included in the PF09 study.}
\tablecomments{This table is available in its entirety in machine-readable form at \url{https://cdsarc.cds.unistra.fr/viz-bin/cat/J/PASP/132/E4301}.}
\end{deluxetable*}

\begin{figure*}[ht]
  \includegraphics[width=0.5\linewidth]{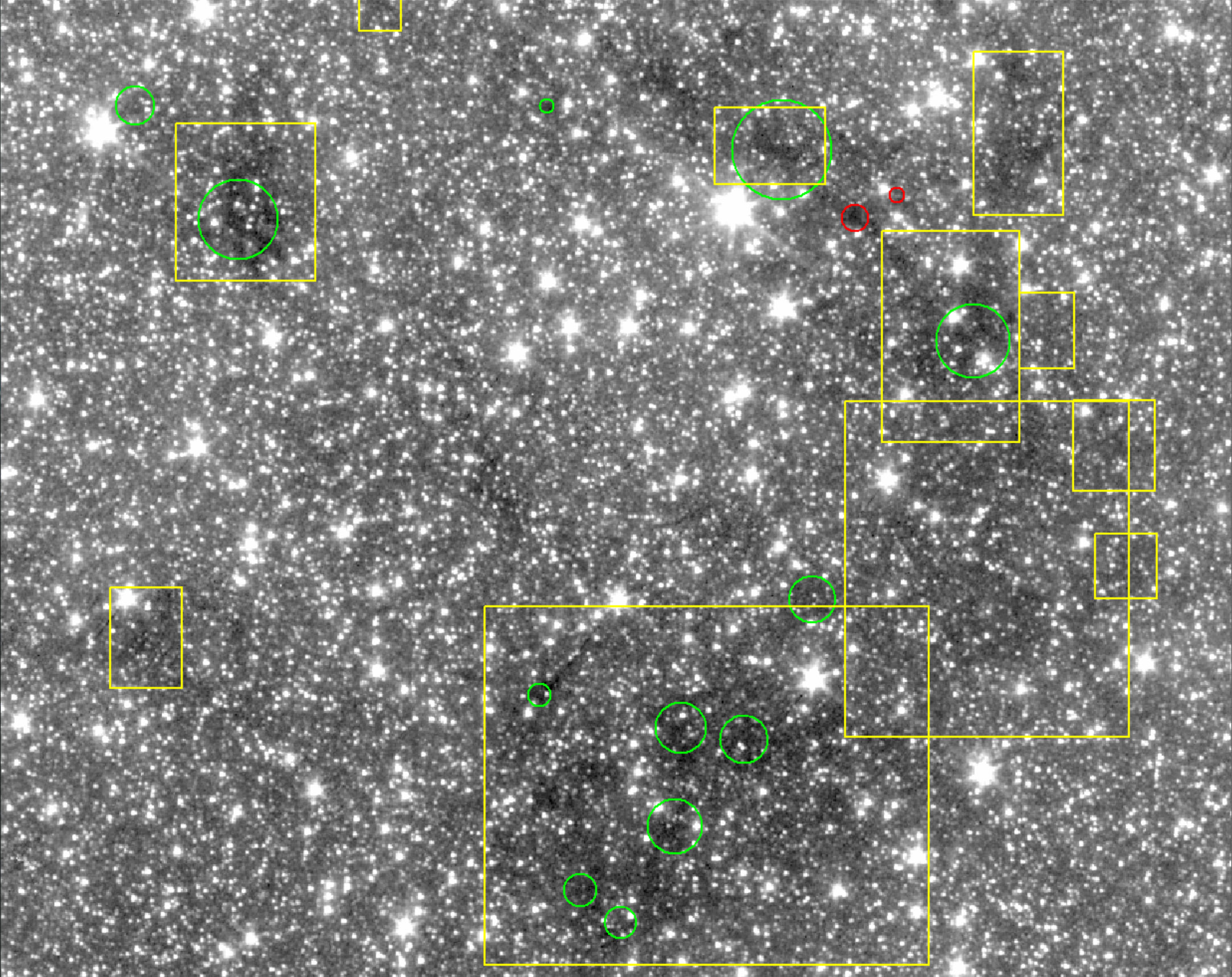}
  \includegraphics[width=0.5\linewidth]{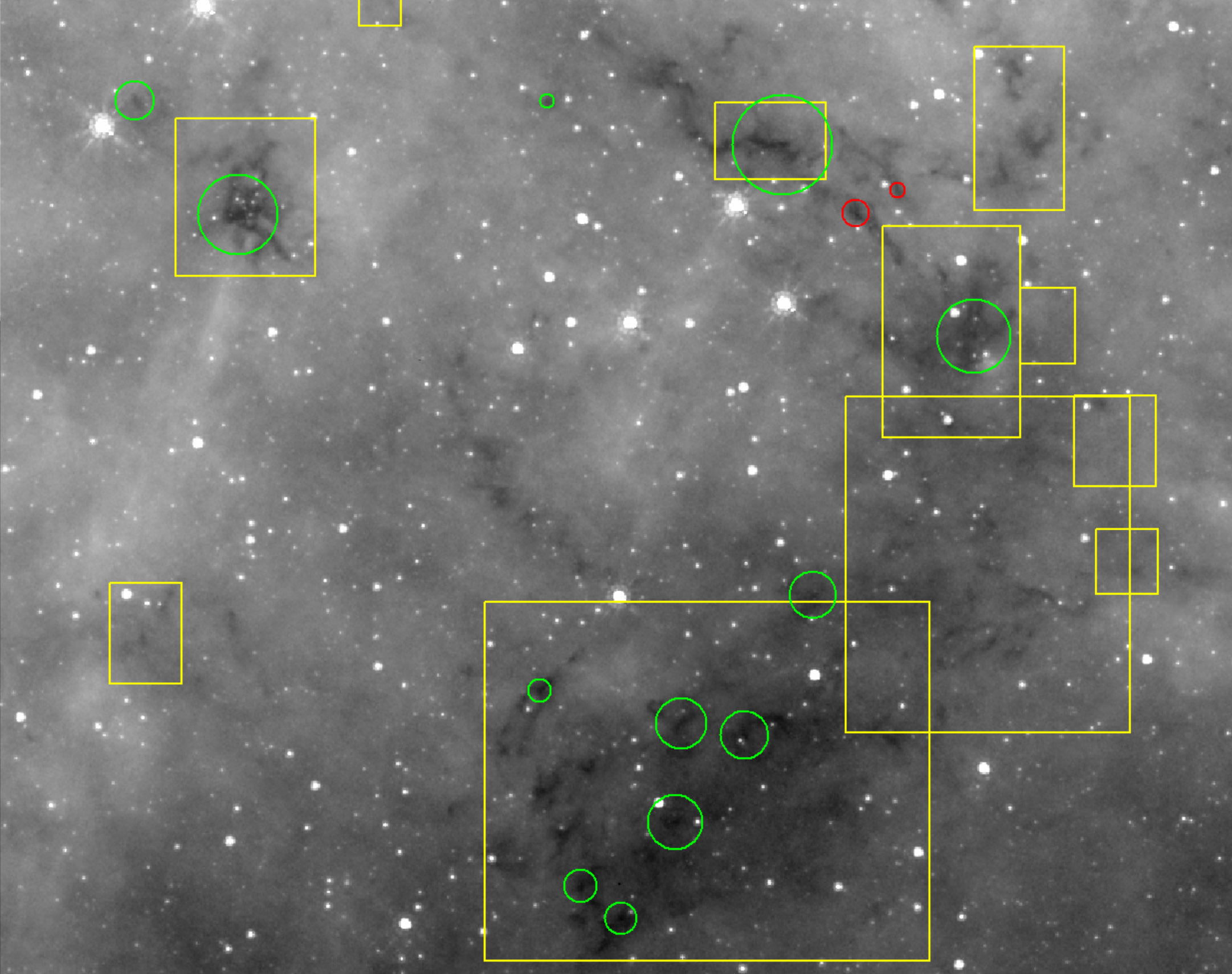}
  \caption{A comparison of the PF09/\cite{Peretto2016} (red and green circles) and our catalog (yellow boxes) in an area covered by both surveys. The region shown is a 17$' \times 13\farcm7$ area centered at $l=29\fdg323$, $b=-0\fdg708$. The 3.6~\micron\ image is on the left, the 8~\micron\ image on the right. The IRDCs found to have a Herschel counterpart by \citet{Peretto2016} are plotted with the green circles with the radius given in that catalog, and the objects with no Herschel source are plotted with red circles. The IRDCs identified in both surveys are largely consistent, with the PF09 catalog tending to identify individual parts of an IRDC as separate objects, whereas our survey outlines larger areas.  Our survey has identified some of the lower density IRDCs that were not marked in the PF09 survey. There appears to be a faint extension of the IRDC just to the left of the center of the image that was missed by both methods.}
  \label{compreg1}
\end{figure*}
\subsection{IRDC Catalog Characteristics and Verification}\label{verify}
To evaluate the accuracy of our technique, we took a sample mosaic from each survey and evaluated by eye the number of true positives, false positives, and false negatives (IRDCs that were missed). A true verification is difficult because there is no other dataset available that would allow us to unambiguously determine whether an IRDC is real or not. It is possible to compare to longer wavelength data such as \citet{Peretto2016} who found Herschel counterparts to the IRDCs in the PF09 catalog. However, that is not a completely reliable method to indicate which objects are true IRDCs, as is evident in the examples shown below. One reason is that many of the small and low-contrast objects found in our survey would not be detected with the factor of 10 or more poorer resolution and often complicated backgrounds in the long wavelength data. We therefore performed a visual test, inspecting by eye each bounding box in the test field to verify whether it contained a real IRDC or not. We used the SAOimageDS9\footnote{\url{http://ds9.si.edu/site/Home.html}} data visualization software to interactively adjust the scale and contrast as we scanned the image in order to best display features at various locations in the image to find IRDCs.  Furthermore, we scanned the image to find any IRDCs that were not identified by the surveys. For the regions where we had access to both the Channel 1 and 4 mosaics, we used both to conduct our inspection because some IRDCs are more visually apparent in the Channel 4 mosaics, and we wanted to evaluate how many objects were possibly being missed because of our use of Channel 1 only in our survey.

The results of our accuracy test for the mosaic GLM\_01200+0000 are shown visually in Figure \ref{Accuracy}, and Table \ref{accuracysummary} summarizes the results for all test regions. We found that our method performed the best in the inner Galaxy, and while the table conveys that some true IRDCs were missed in the test regions, a large fraction of the objects were found and the false positive rates were relatively low. Our method was not able to pick up some faint and small IRDCs, which could be caused by the Gaussian blurring which will tend to reduce our sensitivity to compact and low-contrast objects. This is a trade-off in the survey where the Gaussian blurring improves the IRDC detection and reliability for more distinct objects, but tends to remove some real low-contrast objects.

\par
Most of the IRDCs in our catalog are relatively small, which can be seen in the histogram of apparent IRDC sizes in Figure \ref{Histogram}. The sizes in Table \ref{IRDCcat} and shown in this figure are the projected area within the IRDC contour in units of arcsec$^2$. Without knowing the distances to each IRDC, we cannot determine the true physical sizes of the objects. The distribution of apparent sizes increases fairly smoothly as one moves toward smaller IRDCs, except for a feature in the histogram near an IRDC area of 4200 arcsec$^2$ where the distribution has a slight dip. This does not seem to be related to a change in parameters in the various size ranges discussed in \S \ref{parmchoice} since the feature is fully within the 3000 -- 30,000 size range. The IRDCs below and above this feature do not seem to differ in character in terms of their distribution on the sky or the characteristics of the IRDCs. We therefore do not believe this is a real feature in the IRDC size distribution, but perhaps an artifact of the way that larger IRDCs are broken up into several smaller objects by the algorithm, or perhaps influenced by the way the data are binned for the histogram.

Because most of the objects we find are small IRDCs, most of the false positives tended to also be small, and typically related to small dark spaces between stars in the images. This is because an area with bright stars enclosing a dark area is interpreted as a contour, and its small size makes it hard for the CNN to classify it accurately. Many of the false positives were removed using a lower limit cutoff (see \S \ref{method}), but the cutoff was not set high enough to remove all of them since that would have removed many true IRDCs as well. In practice, any IRDC survey will have an inherent lower size cutoff due to the instrument resolution, sensitivity, and characteristics of the background mission. Here we have explicitly set that value slightly higher than the theoretical limit, at the cost of removing some true IRDCs that are smaller than the cutoff.

We noticed that the performance in the Cygnus-X region was poorest in terms of fraction of false positives and false negatives compared to the number of true IRDCs in this region. Many of the false positives in the Cygnus-X region tend to be low emission regions surrounded by brighter emission features. These structures mimic the appearance of IRDCs, but are often more diffuse and not as centrally concentrated as the true IRDCs. There were also a significant fraction of missed IRDCs in the Cygnus-X region. 
These issues could be due to the fact that our classifier was trained on the inner parts of the Galaxy, where the IRDCs are more distant (typically 3-5~kpc compared to the ones in Cygnus-X which are 1.4~kpc or less) and there was insufficient data with low emission regions to train the classifier. This would be an issue that could possibly be addressed through transfer learning, where our model weights could be used as a starting point to train on a new dataset. However, we chose not to develop new methods for this region, and present the catalog that used uniform techniques over the entire dataset.\label{cygnusX}

\subsection{Comparison to the PF09 and MSX Catalogs}
We compared our catalog to the IRDC catalogs from PF09 along with the MSX catalog by \cite{Simon_2006}. The PF09 catalog provides an estimate of the size of the IRDC in two-dimensions and the orientation of the major axis, and Simon et al. provide the major and minor axes sizes and position angle of the major axis, whereas we provide rectangles to define the regions occupied by IRDCs. Therefore, for a long and thin IRDC that runs diagonally in Galactic coordinates, the PF09 and MSX catalogs will have a better representation of an IRDC's dimensions and orientation than our bounding boxes. However, from the example images shown here, one can see that many IRDCs are irregular and there is no clear orientation. Our method as well as PF09 tend to split large IRDCs into smaller segments, so the sizes we find do not vastly misrepresent the area of the IRDC. In addition, in Table \ref{IRDCcat} we give the contour area, which is a better estimate of the IRDC area than the bounding box or an estimate of the major and minor axes of the IRDC.

Both of the previous catalogs also used computational approaches for IRDC localization, and in Figure \ref{Overlay} we show a sample region where we overlay these catalogs along with ours. As one would expect from the lower spatial resolution of the MSX survey, the IRDCs in that catalog are larger, and tend to enclose regions that are broken up into individual IRDCs by the other surveys. There are some false positives that seem to be due to regions of lower emission surrounded by brighter regions, rather than absorption of emission by an IRDC. This is also an issue for the PF09 (see below) and our survey (see \S \ref{cygnusX}). The orientation and aspect ratio of the IRDCs seem to track the actual shapes reasonably well, although in many cases they seem to overestimate the IRDC size. 

The PF09 catalog and ours tend to break up larger IRDCs and ones with complicated structure into smaller objects. This characteristic of the search technique is sensitive to the threshold level set, which can break up an object into different contours and then be identified as separate objects, whereas visually they appear to be parts of a larger spatially complex IRDC. The PF09 catalog seemed to further break apart IRDCs compared to ours. Some of the larger and more irregular IRDCs are better represented in terms of the cloud area and location in our survey whereas in the PF09 catalog only small parts of the clouds were identified.  It can be seen in Figure \ref{Overlay} that many of the blue boxes of our survey contain many small red PF09 objects. In this field, there were no PF09 IRDCs larger than the matching object in our catalog. In the full catalog, we found a total of 12331 IRDCs in the regions covered by the PF09 catalog. Of these, 2626 of our IRDCs had one or more PF09 objects with centers within the IRDC boxes listed in our catalog. A total of 4102 PF09 objects were within our IRDC regions. Regions in the two catalogs which did not overlap are a mix of PF09 objects that identify parts of IRDCs that we also found but their centers lie outside of our bounding boxes; IRDCs that are below our lower size cutoff (5769 objects) or otherwise rejected from our catalog; and possible false positives in the catalogs. For the PF09 objects below our size cutoff, many of these lie within the larger IRDC regions identified in our survey, as is apparent in Figure \ref{Overlay}, so they do not all correspond to IRDCs that are missing from our survey.
\par

\begin{deluxetable*}{ccccc}
\caption{Summary of Accuracy Tests}\label{accuracysummary}
\tablehead{\colhead{Mosaic} &\colhead{Survey}& \colhead{False Positives} & \colhead{True Positives} & \colhead{False Negatives}}
\startdata
GLM\_01200+0000 & I& 1 & 456 & 43 \\ 
GLM\_00000+0000 & II & 0 & 438 & 48 \\
GLM\_00000+0300 & 3D & 0 &5 & 3\\
corr\_CYGX\_07800+0250& CygX & 11 & 137 & 31\\
corr\_GLM\_02700+0150 & Deep& 4 & 209 & 9 \\
corr\_GLM\_06600+0105 & 360 & 32 & 166 & 0 \\
corr\_SMOG\_10500+0145 & SMOG & 7 & 0 & 1 \\
corr\_VELACAR\_25800+0050& VelaCar & 1 & 0 & 0 \\
\enddata
\end{deluxetable*}

\begin{figure*}[ht]
\begin{center}
  \includegraphics[width=0.45\linewidth]{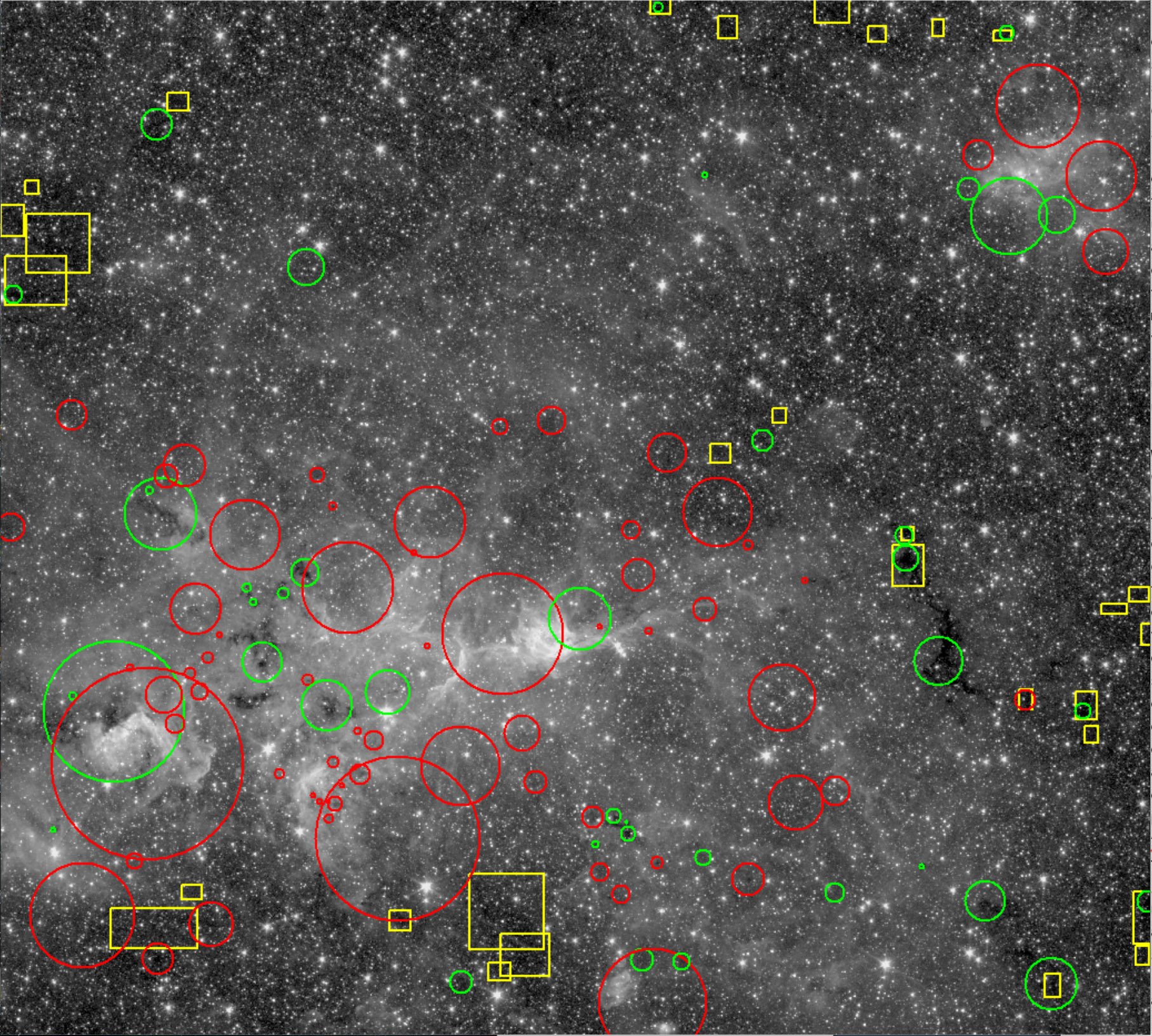}
  \includegraphics[width=0.44775\linewidth]{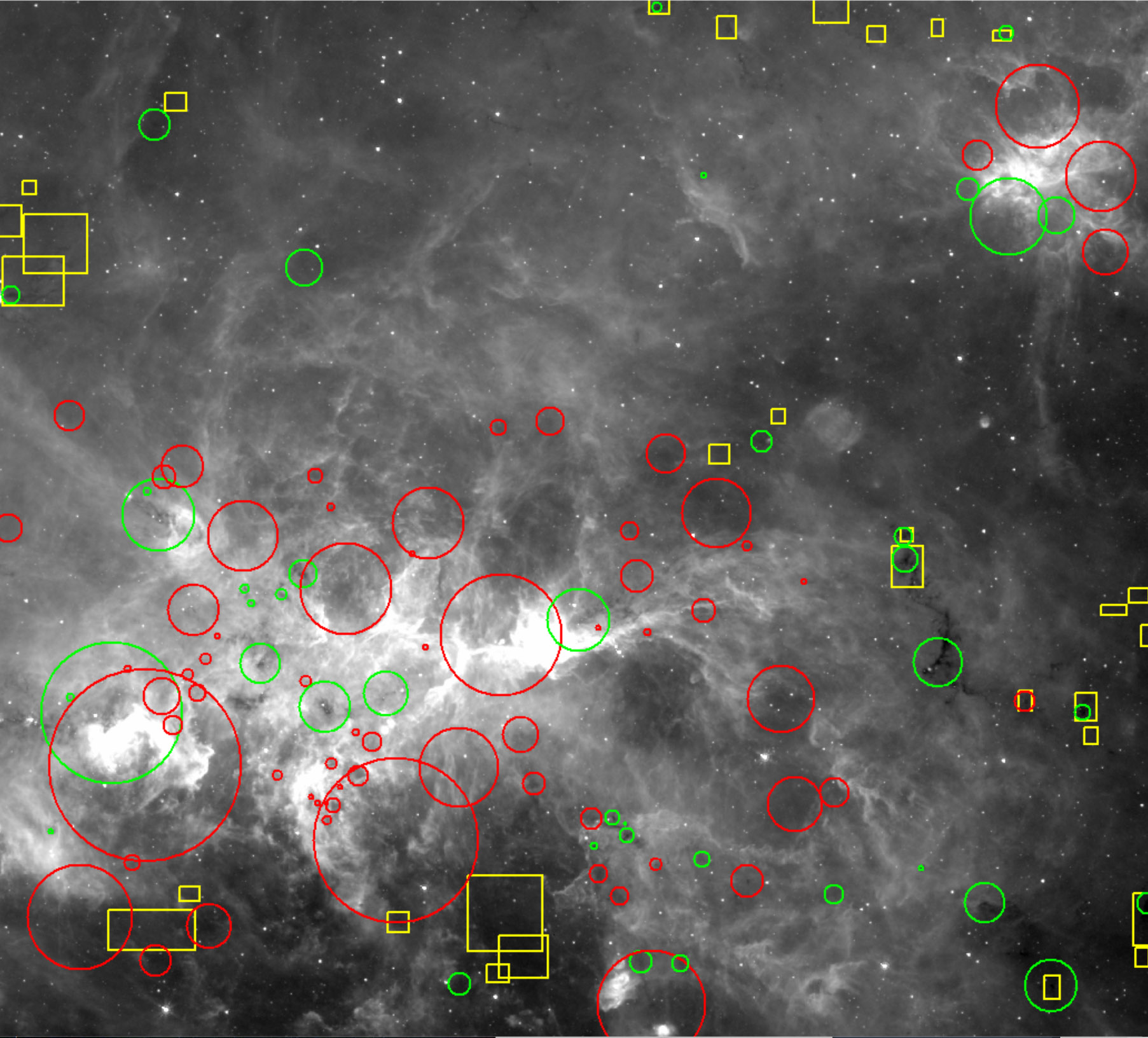}
\end{center}
  \caption{A comparison of the PF09 and our catalog, using the same symbols as in Figure \ref{compreg1}. The region shown is a 46$' \times 41'$ area centered at $l=48.905$\degr, $b=-0.239$\degr. The 3.6~\micron\ image is on the left, the 8~\micron\ image on the right. There is some correspondence between the surveys in the lower emission regions, but near the regions with bright 8~\micron\ emission the PF09 survey has identified many regions that appear to be lower intensity cavities near bright emission features. Many, but not all, of these regions were flagged by \citet{Peretto2016} as not having Herschel counterparts (plotted in red). Our survey seems to have avoided identifying objects in the bright areas, missing a few true IRDCs that appear in the lower left part of the region.}
  \label{compreg2}
\end{figure*}

\par
An example comparing the performance of the two surveys against the 3.6 and 8~\micron\ images of a field is shown in Figure \ref{compreg1}. The PF09 (green and red) and our catalog objects (yellow) are plotted on both images of the same region. The PF09 objects found to have a Herschel counterpart by \citet{Peretto2016} are plotted with green circles, the objects without are plotted in red. In general there is good correspondence between objects in the two catalogs in the region shown. The IRDCs identified in both surveys are largely consistent, with the PF09 catalog tending to identify individual parts of an IRDC as separate objects, whereas our survey outlines larger areas. This is by design since we have removed smaller regions identified within larger IRDCs. There are a few IRDCs that we found that were not listed in PF09, perhaps due to their low contrast with surrounding regions, and a couple IRDCs in the PF09 catalog that we did not list, due to their small size. The PF09 objects in this region without Herschel counterparts appear to be true IRDCs. There is a faint extension of the IRDC in the center of the image that was not identified in either survey.

A further comparison is shown in Figure \ref{compreg2}, which shows a larger region with some areas of bright emission in both channels. Again the 3.6 and 8~\micron\ images of the same region are shown and the PF09 objects overlaid in green and red, and our survey in yellow. Here there is lower correspondence between the two surveys, with the majority of PF09 sources being related to the bright emission features in the region. A large number of those catalog entries seem to be related to areas of lower emission surrounded by bright features, and are likely not true IRDCs. Many of these PF09 objects do not have Herschel counterparts (red circles). There are several objects that appear to be true IRDCs located in those bright regions that our survey failed to locate, probably due to the complicated backgrounds in the region. There is better agreement between the surveys in the lower background regions around the edges of the image. 

\begin{figure}[ht]
  \vskip 0.2in
  \includegraphics[width=\linewidth]{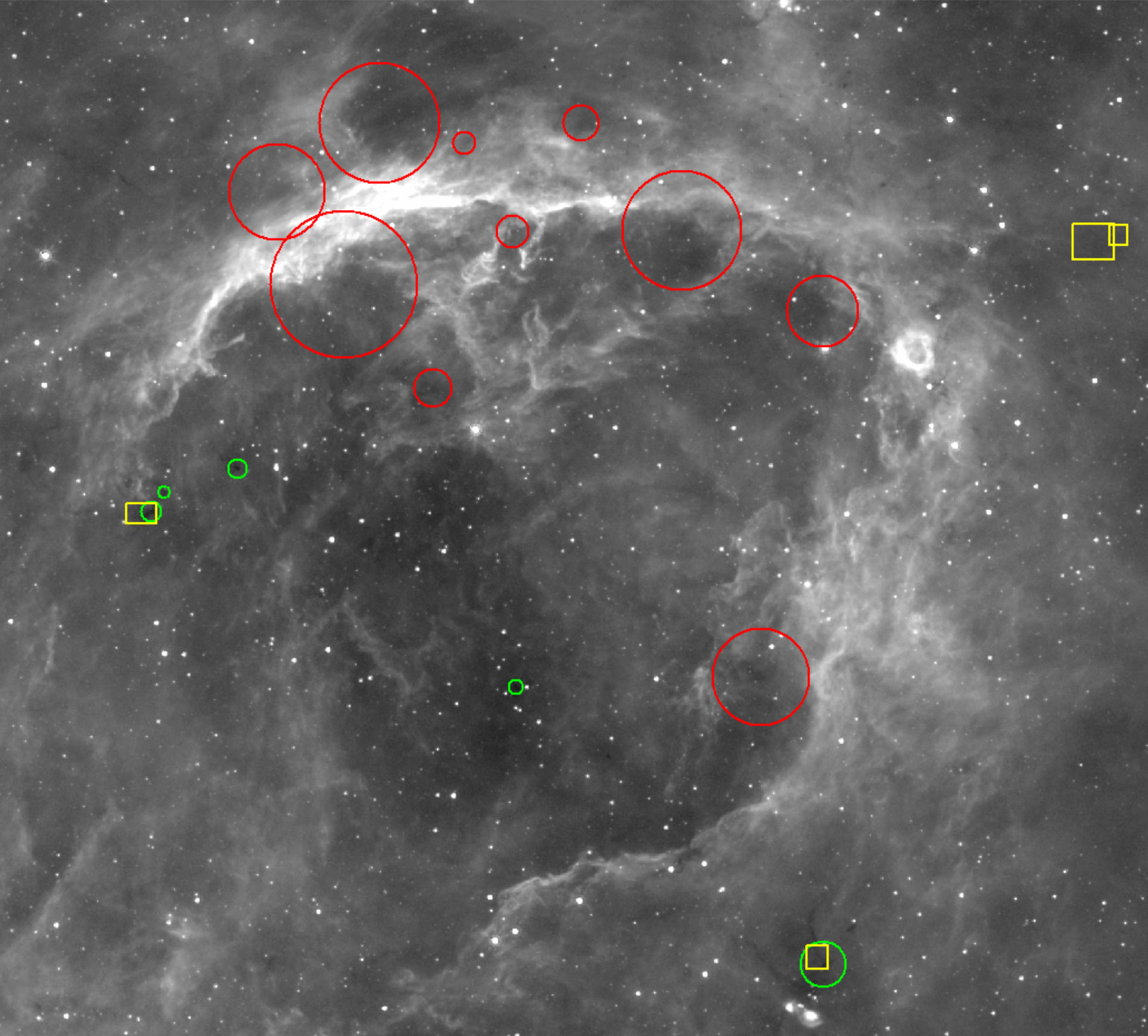}
  \caption{A comparison of the PF09 and our catalog, using the same symbols as in Figure \ref{compreg1}. The region shown is a $34\arcmin \times 29\arcmin$ area of the 8~\micron\ mosaic centered at $l=50\fdg943$, $b=0\fdg091$. The PF09 survey has identified several large regions near the bright-rimmed cloud as IRDCs where they appear to be regions of lower emission near the bright filamentary structure. These are plotted with the red circles, indicating that \citet{Peretto2016} found these to not have Herschel counterparts. Several smaller objects that appear to be true IRDCs were located by PF09 (green circles) and our survey (yellow) in the region as well.}
  \label{BRR1}
\end{figure}

Figure \ref{BRR1} shows another example of objects identified in the surveys near a bright-rimmed cloud. Many of the PF09 objects seem to be related to the emission voids near the bright emission regions. Fifteen objects are listed in the PF09 survey and three in our survey. One object is common to both surveys, and ten of the larger regions drawn in red were found by \citet[][]{Peretto2016} not to have Herschel counterparts. The five PF09 objects drawn in green locate what appear to be true small IRDCs, two of which our survey also found. The algorithm used in our survey largely ignored the areas near the bright rimmed cloud and found two low contrast regions near the right edge of the image. 

\begin{figure}[ht]
  \includegraphics[width=\linewidth]{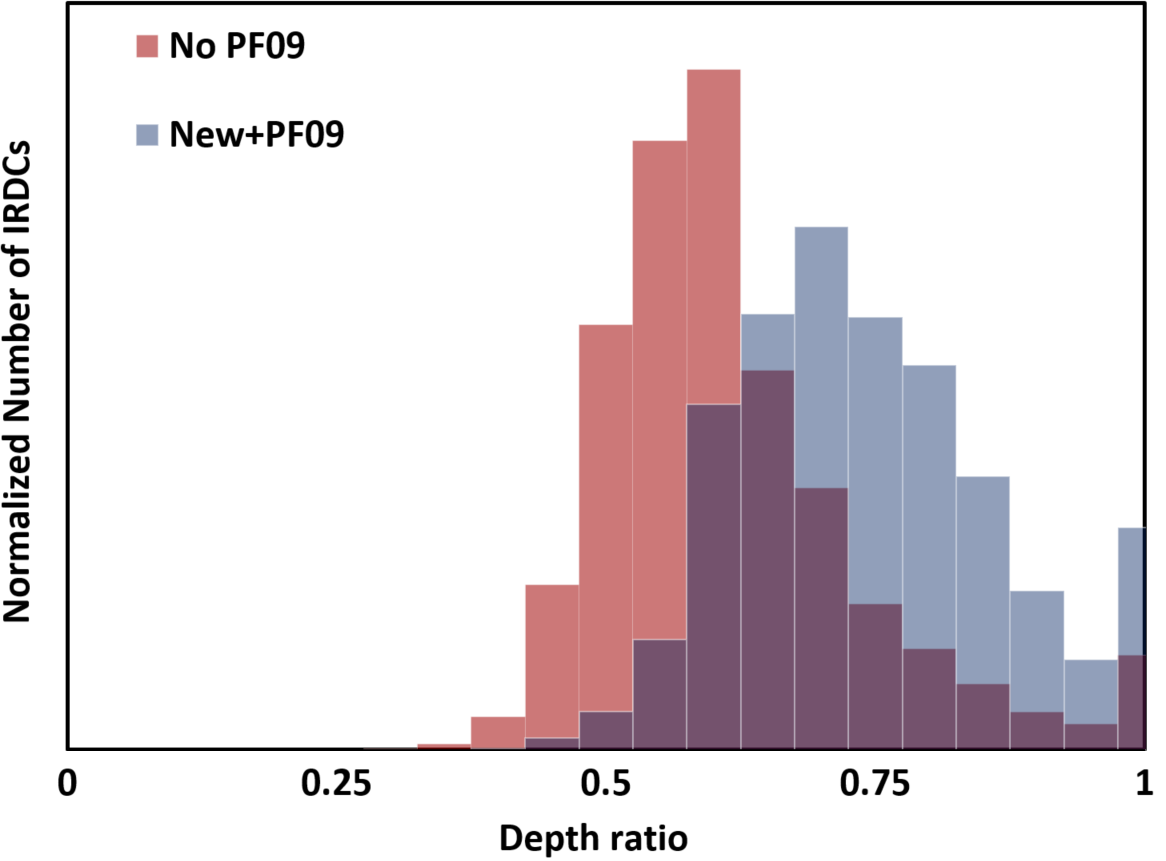}
  \caption{Histograms of the IRDC depth ratios as measured by our survey in the 3.6~\micron\ images. The blue histogram shows IRDCs that are in both our new and the PF09 surveys, the pink histogram shows objects in our survey that are in the PF09 area but not found by that survey (where the histograms overlap is shown in purple). The vertical plot ranges were normalized to the same fraction of the total number of objects in that histogram. Objects common to both surveys have higher depth ratios than regions found only in our survey. }
  \label{densityhist}
\end{figure}

Our survey located a number of objects with low density that perhaps did not satisfy the density cutoff used by PF09. Figure \ref{densityhist} shows histograms of the IRDC depth ratios as measured by our survey in the 3.6~\micron\ images. The blue histogram shows IRDCs that are in both surveys, the pink histogram shows objects in our survey that are in the PF09 area but not listed in that survey. The vertical plot ranges were normalized to the same fraction of the total number of objects in that histogram. The pink histogram is more sharply peaked and has a median value (0.57) lower than the histogram of objects in both surveys (0.71), indicating that objects common to both surveys have on average higher depth ratios than regions found only in our survey. 

\begin{figure*}[ht]
\begin{center}
\vskip 0.2in
  \includegraphics[width=5in]{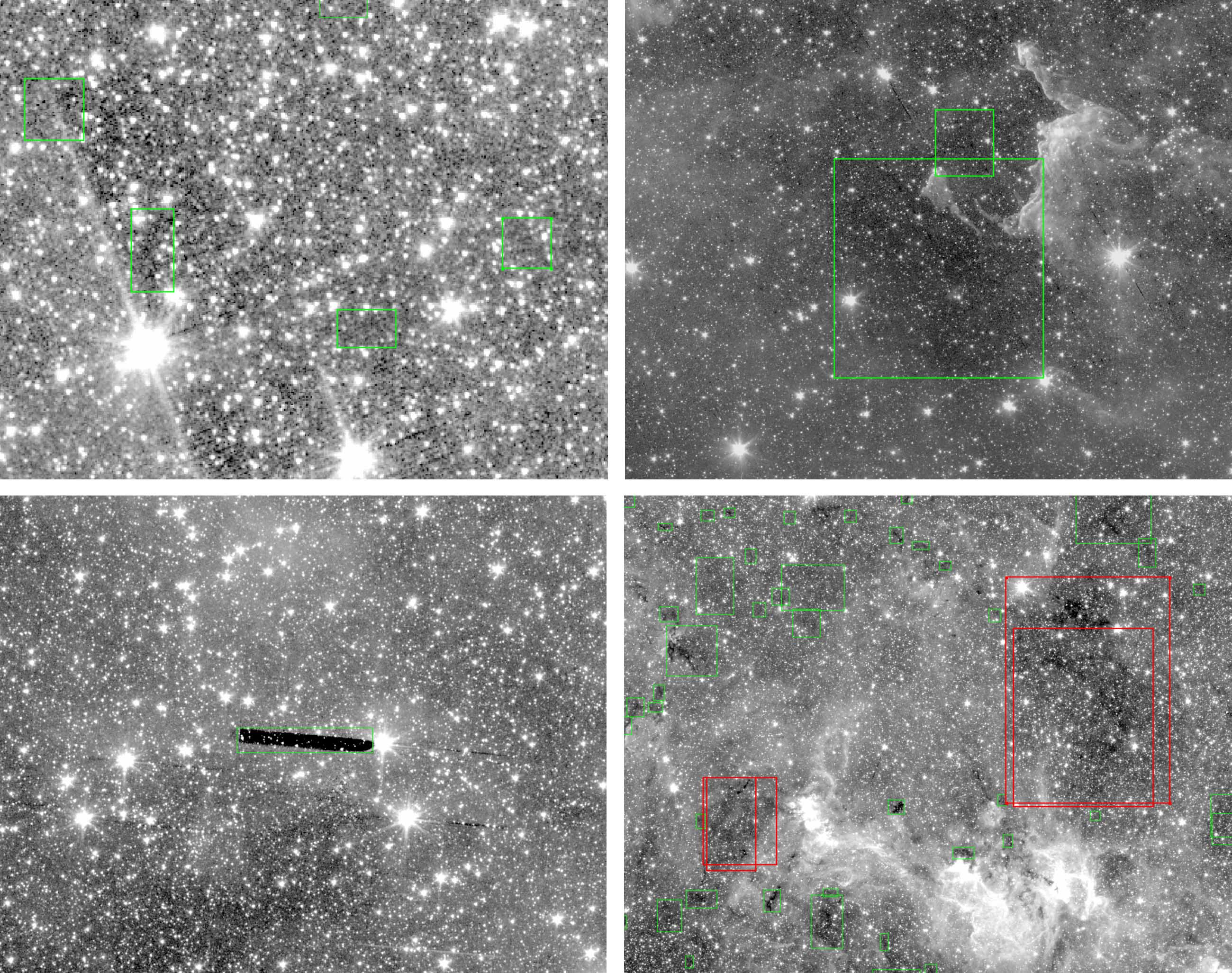}
  \caption{Examples of some issues with the images and IRDC identifications in our catalog. See \S \ref{limitations} for a description of these cases. Top Left: The region near G295.58--01.011 which a bright star causes artifacts in the image and the region is misidentified as an IRDC. Top Right: The large region G172.86--00.455. Lower Left: G263.00--01.111 which is caused by an uncorrected pulldown effect in the original IRAC images. Lower Right: The region including G332.98--00.221, which shows two IRDC regions (red rectangles) that are largely overlapping with other objects in the catalog.}
  \label{ErrExamples}
\end{center}
\end{figure*}
In summary, our survey is largely consistent with and in some ways complementary to the PF09 catalog. For low background and uncomplicated regions, the surveys often identify the same objects, with the PF09 highlighting the densest individual components and our survey grouping them into single larger IRDCs. The PF09 catalog may contain small dense IRDCs that ours will miss because of the lower size cutoff, and our survey appears to be more sensitive to low-contrast regions that might be below the minimum column density requirement of the PF09 catalog. In regions of bright emission, the PF09 catalog seems more prone to including emission voids near bright features, but also locates some true IRDCs in these regions that our survey missed because of the complicated background. The \citet{Peretto2016} comparison to the Herschel sources helps to indicate which PF09 IRDCs might be emission voids in the 8~\micron\ images rather than absorption due to an IRDC, but there are also apparently true IRDCs that are not verified in the Herschel data due to their size or the complex nature of some regions.

\subsection{Limitations of the IRDC Catalog}\label{limitations}
There are several limitations to the IRDC catalog presented here, some of which are described in \S \ref{verify} above. Figure \ref{ErrExamples} shows some examples of problem areas in the survey images. The region G295.58$-$01.011 was identified near a bright star because of an uncorrected ``muxbleed'' effect \citep[see][for a description of this and other IRAC array artifacts]{2004SPIE.5487...77H} and another bright source artifact where the background level of the array is artificially depressed in certain rows of the array near a bright star. This causes the algorithm to detect two IRDCs along the region near the bright star. Another region shown in Figure \ref{ErrExamples} is the large IRDC G172.86$-$00.455. There does appear to be a true IRDC in the center of this region, but it is also in a void of low emission surrounded by higher diffuse emission and a bright-rimmed cloud along the right side of the area. The algorithm was misled by this structure and overestimated the extent of the actual IRDC. In the lower left panel of Figure \ref{ErrExamples}, the catalog object G263.00$-$01.111 locates a column pulldown artifact near a bright star in the mosaic. Finally, in the lower right panel, G332.98$-$00.221 shows an example of two IRDC regions (in red) which were each identified as two separate IRDCs. Although they are mostly overlapping, there was enough of a difference between their centers and coverage areas that they were not removed as duplicates in our search for these objects described above.

Some IRDCs will be missed because they are too low contrast and/or too small to be identified, given the survey sensitivity and resolution of \Sp, or because of the characteristics of the local background emission. An IRDC might have low contrast because it does not contain much absorbing mass, or it might not be in front of sufficiently bright background emission for it to stand out in the image. When the stellar density and background level is low, gaps between stars can be mistaken for regions of high absorption. This has been partially mitigated by using a lower limit on the IRDC size, but this will not remove all of these spurious IRDCs, and it removes the smallest true IRDCs from the list that fall below the cutoff size.

Bright emission and spatially complicated background regions can lead to missed IRDCs, or false positives when a lower emission region is surrounded by bright source structure. Large individual IRDCs or complexes are often broken up by this algorithm into separate objects, and separate IRDCs can be perceived as one object if they overlap spatially, even though they could be physically distant from each other. Another possibility is that an IRDC might have a unique morphology that is not recognized by the algorithm as an IRDC because it is not well-represented by the objects in the training set or the other members of the catalog. Most of these limitations exist to some degree for all search techniques, so any catalog can never be complete or accurately characterize all IRDCs. However, our method performs well when compared to the previous search methods used, and appears to find a large fraction of true IRDCs while minimizing false positives. It also avoids breaking up IRDCs into many individual objects, which would affect the perceived distribution of IRDC sizes.

\section{Conclusions}
Our goal was to use an objective, automated method for finding IRDCs in the large GLIMPSE dataset that could be applied to the regions where only \Sp\ Warm Mission (3.6 and 4.5~\micron) data are available. We used open source astronomical, image processing, and neural network libraries to develop a method that sought to maximize the identification of true IRDCs and minimize the number of false positives and false negatives. We applied this method to the various GLIMPSE and other Galactic plane datasets to create an IRDC catalog over the full range of Galactic longitude covered in these surveys. The catalog identifies many of the previously known IRDCs found in the earlier MSX and PF09 surveys, and finds new objects in survey regions that have recently become available late in the \Sp\ mission.

\acknowledgements
This work is based on observations made with the \Sp\ Space Telescope, which is operated by the Jet Propulsion Laboratory, California Institute of Technology under a contract with NASA. Support for this work was provided by NASA through an award issued by JPL/Caltech. We also thank artist Mario Klingemann who was very helpful when dealing with the Tensorflow aspect of our procedure. 
\facility{Spitzer (IRAC)}
\software{Astropy\citep{astropy}, Pillow, OpenCV, Tensorflow, SAOimageDS9}
\bibliographystyle{aasjournal}
\bibliography{pari_biblio}

\end{document}